\documentclass[longauth]{aa} % for the long lists of affiliations 
\usepackage{graphicx}
\usepackage{ulem}
\usepackage[varg]{txfonts}
\usepackage{natbib}
\usepackage{multirow}

\usepackage{enumerate}
\usepackage{booktabs}

\newcommand{\mH}{\ensuremath{{m_{\rm H}}}}
\newcommand{\nH}{\ensuremath{{n_{\rm H}}}}

\newcommand{\Av}{\ensuremath{A_{\rm V}}}

\newcommand{\CO}{\ensuremath{^{12}\mathrm{CO}}}
\newcommand{\tCO}{\ensuremath{^{13}\mathrm{CO}}}
\newcommand{\CeO}{\ensuremath{\mathrm{C^{18}O}}}
\newcommand{\CsO}{\ensuremath{\mathrm{C^{17}O}}}
\newcommand{\CtS}{\ensuremath{\mathrm{C^{34}S}}}
\newcommand{\NtHp}{\ensuremath{\mathrm{N_{2}H^{+}}}}
\newcommand{\Tdust}{\ensuremath{T_{\rm dust}}}
\newcommand{\NHH}{\ensuremath{N({\rm H}_2)}}

\newcommand{\Tmb}{\ensuremath{ T_{\rm mb} }}
\newcommand{\Tex}{\ensuremath{ T_{\rm ex} }}

\newcommand{\Vlsr}{\ensuremath{ v_{\rm lsr} }}

\newcommand{\HII}{H{\sc II}}

\begin{document}

   \title{Multi-scale analysis of the Monoceros OB 1 star-forming region
   \thanks{The reduced datacubes (FITS files) of our IRAM and TRAO observations
   are only available at the CDS via anonymous ftp to cdsarc.u-strasbg.fr 
   (130.79.128.5) or via 
   http://cdsarc.u-strasbg.fr/viz-bin/qcat?J/A+A/???/A??}}
   \subtitle{II. Colliding filaments in the Monoceros OB1 molecular cloud}
   \author{Julien Montillaud\inst{1}
         \and
         Mika Juvela\inst{2}
         \and
         Charlotte Vastel\inst{3}
         \and
         J.H. He\inst{4,5,6} %\fnmsep\thanks{ORCID 0000-0002-3938-4393}
         \and
         Tie Liu\inst{7,8,9}
         \and
         Isabelle Ristorcelli\inst{3}
         \and
         David J. Eden\inst{10}
         \and
         Sung-ju Kang\inst{8}
         \and
         Kee-Tae Kim\inst{8,11}
         \and
         Patrick M. Koch\inst{12}
         \and
         Chang Won Lee\inst{8,11}
         \and
         Mark G. Rawlings\inst{9}
         \and
         Mika Saajasto\inst{2}
         \and
         Patricio Sanhueza\inst{13}
         \and
         Archana Soam\inst{14,8}
         \and
         Sarolta Zahorecz\inst{15,13}
         \and
         Dana Alina\inst{16}
         \and
         Rebeka Bögner\inst{1,17}
         \and
         David Cornu\inst{1}
         \and
         Yasuo Doi\inst{18}
         \and
         Johanna Malinen\inst{19}
         \and
         Douglas Marshall\inst{20}
         \and
         E. R. Micelotta\inst{2}
         \and
         V.M. Pelkonen\inst{1,2,21}
         \and
         L. V. Tóth\inst{17,22}
         \and
         A. Traficante\inst{23}
         \and
         Ke Wang\inst{24,25}
         }

% J. Montillaud, M. Juvela, C. Vastel, J.H. He, Tie Liu, I. Ristorcelli, D. J. Eden, Sung-ju Kang, Kee-Tae Kim, P. M. Koch, Chang Won Lee, M. G. Rawlings, M. Saajasto, P. Sanhueza, A. Soam, S. Zahorecz, D. Alina, R. Bögner, D. Cornu, Yasuo Doi, et al.
% J. Malinen, D. Marshall, E. R. Micelotta, V.M. Pelkonen, L. V. Tóth, A. Traficante, Ke Wang

   \institute{
      Institut UTINAM - UMR 6213 - CNRS - Univ Bourgogne Franche Comté, France, OSU THETA, 41bis avenue de l'Observatoire, 25000 Besan\c{c}on, France - e-mail: {\tt julien@obs-besancon.fr}
      \and %2
      Department of Physics, P.O.Box 64, FI-00014, University of Helsinki, Finland
      \and %3
      IRAP, Université de Toulouse, CNRS, UPS, CNES, F-31400 Toulouse, France
      \and %4
      Yunnan Observatories, Chinese Academy of Sciences, 396 Yangfangwang, Guandu District, Kunming, 650216, P. R. China
      \and %5
      Chinese Academy of Sciences, South America Center for Astrophysics (CASSACA), Camino El Observatorio 1515, Las Condes, Santiago, Chile
      \and %6
      Departamento de Astronom\'ia, Universidad de Chile, Las Condes, Santiago, Chile
      \and %7
      Shanghai Astronomical Observatory, Chinese Academy of Sciences, 80 Nandan Road, Shanghai 200030, China
      \and %8
      Korea Astronomy and Space Science Institute, 776 Daedeokdaero, Yuseong-gu, Daejeon 34055, Republic of Korea
      \and %9
      East Asian Observatory, 660 N. A’ohoku Place, Hilo, HI 96720, USA
      \and %10
      Astrophysics Research Institute, Liverpool John Moores University, IC2, Liverpool Science Park, 146 Brownlow Hill, Liverpool L3 5RF, UK
      \and %11
      University of Science \& Technology, 176 Gajeong-dong, Yuseong-gu, Daejeon, Republic of Korea
      \and % 12
      Academia Sinica, Institute of Astronomy and Astrophysics, Taipei, Taiwan
      \and % 13
      National Astronomical Observatory of Japan, National Institutes of Natural Sciences, 2-21-1 Osawa, Mitaka, Tokyo 181-8588, Japan
      \and % 14
      SOFIA Science Centre, USRA, NASA Ames Research Centre, MS N232-12 Moffett Field, CA 94035, USA
      \and % 15
      Department of Physical Science, Graduate School of Science, Osaka Prefecture University, 1-1 Gakuen-cho, Naka-ku, Sakai, Osaka 599-8531, Japan
      \and % 16
      Department of Physics, School of Science and Humanities, Kabanbay batyr ave, 53, Nur-Sultan 010000 Kazakhstan
      \and %17
      Eötvös Loránd University, Department of Astronomy, Pázmány Péter sétány 1/A, H-1117, Budapest, Hungary
      \and % 18
      Department of Earth Science and Astronomy, Graduate School of Arts and Sciences, The University of Tokyo, 3-8-1 Komaba, Meguro, Tokyo 153-8902, Japan
      \and % 19
      Institute of Physics I, University of Cologne, Zülpicher Str. 77, D-50937, Cologne, Germany
      \and % 20
      AIM, CEA, CNRS, Université Paris-Saclay, Université Paris Diderot, Sorbonne Paris Cité, F-91191 Gif-sur-Yvette, France
      \and % 21
      ICC, University of Barcelona, Marti i Franquès 1, E-08028 Barcelona, Spain
      \and % 22
      Konkoly Observatory of the Hungarian Academy of Sciences, H-1121 Budapest, Konkoly Thege Miklósút 15-17, Hungary
      \and % 23
      IAPS-INAF, via Fosso del Cavaliere 100, I-00133, Rome, Italy
      \and %24
      Kavli Institute for Astronomy and Astrophysics, Peking University, 5 Yiheyuan Road, Haidian District, Beijing 100871, China
      \and %25
      European Southern Observatory (ESO) Headquarters, Karl-Schwarzschild-Str. 2, 85748 Garching bei M\"{u}nchen, Germany
      }

\date{Received 18 December 2018; Accepted 26 August 2019}
% \abstract{}{}{}{}{} 
% 5 {} token are mandatory
   \abstract
   { We started a multi-scale analysis of star formation in G202.3+2.5, an 
   intertwined filamentary sub-region of the Monoceros OB1 molecular complex, in
   order to provide observational constraints on current theories and models that
   attempt to explain star formation globally. In the first paper (Paper I), we 
   examined the distributions of dense cores and protostars and found enhanced 
   star formation activity in the junction region of the filaments.}
   { In this second paper, we aim to unveil the connections between the core 
   and filament evolutions, and between the filament dynamics and the global 
   evolution of the cloud.}
   { We characterise the gas dynamics and energy balance in different parts
   of G202.3+2.5 using infrared observations from the {\it Herschel} and WISE
   telescopes and molecular tracers observed with the IRAM 30-m and TRAO 14-m 
   telescopes. The velocity field of the cloud is examined and velocity-coherent
   structures are identified, characterised, and put in perspective with the 
   cloud environment.}
   { Two main velocity components are revealed, well separated in radial 
   velocities in the north and merged around the location of intense \NtHp\, 
   emission in the centre of G202.3+2.5 where Paper I found the peak of star 
   formation activity. We show that the relative position
   of the two components along the sightline, and the velocity gradient of the
   \NtHp\, emission imply that the components have been undergoing collision for
   $\sim 10^5$ yrs, although it remains unclear whether the gas moves mainly 
   along or across the filament axes. The dense gas where \NtHp\, is detected 
   is interpreted as the compressed region between the two filaments, which 
   corresponds to a high mass inflow rate of $\sim 1\times 10^{-3}\,M_\odot/$yr 
   and possibly leads to a significant increase in its star formation efficiency.
   We identify a protostellar source in the junction region that possibly powers 
   two crossed intermittent outflows.
   We show that the \HII\, region around the nearby cluster NCG 2264 is still
   expanding and its role in the collision is examined. However, we cannot rule
   out the idea that the collision arises mostly from the global collapse of the
   cloud.
   }
   { The (sub-)filament-scale observables examined in this paper reveal
   a collision between G202.3+2.5 sub-structures and its probable role in feeding
   the cores in the junction region. To shed more light on this link between 
   core and filament evolutions, one must characterise the cloud morphology, its 
   fragmentation, and magnetic field, all at high resolution. We consider the
   role of the environment in this paper, but a larger-scale study of this
   region is now necessary to investigate the scenario of a global cloud collapse.
   }


   \maketitle
%
%________________________________________________________________

%=======================================================================================================================
% Introduction

\section{Introduction}  \label{sec:introduction}

%   \subsection{Multiscale view of star formation}

   Understanding star formation is a central challenge in astrophysics, impacting
   the physics and chemistry of the interstellar medium (ISM), stellar physics, 
   as well as the evolution of galaxies and their stellar populations. A wealth
   of studies have been conducted to characterise star formation from the 
   sub-parsec scales (where the ultimate gravitational instability turns prestellar
   cores into protostars) to galactic scales (where the large-scale dynamics 
   shapes the distribution of giant molecular clouds) by way of the molecular 
   cloud scale whose dynamics give birth to filaments and cores. At all scales
   the challenge is to untangle the complex interplay between gravity, turbulence,
   and the magnetic field. Additionally, the stakes are raised by the fact that various
   scales are coupled through non-linear processes, such as dynamical instabilities,
   supernovae explosions or ionisation, and photodissociation fronts.

   Impressive progress has been made in the last few decades in the spatial
   resolution of star formation simulations, enabling the modelling of several
   orders of magnitude of physical scales in a single run. \citet{renaud_sub-parsec_2013}
   modelled the hydrodynamics of a full Milky Way-like galaxy down to 0.05 pc, 
   revealing how the spiral arms pump turbulent energy into the gas and determine 
   the space and mass distributions of molecular clouds. They conclude that 
   gravitation can govern the hierarchical organisation of structures from the 
   galactic scale down to a few parsecs. This is in line with the 'hierarchical 
   gravitational collapse', a scenario \citet{vazquez-semadeni_high-_2009, 
   vazquez-semadeni_global_2019} propose in which small-scale infall motions occur within 
   large-scale ones, driving the dynamics and morphology of star formation regions. 
   This dynamic view of a global collapse is particularly invoked for the formation
   process of high-mass stars, as supported by a number of observational studies (\citeauthor
   {schneider_dynamic_2010} \citeyear{schneider_dynamic_2010}, \citeauthor
   {csengeri_gas_2011} \citeyear{csengeri_gas_2011}, \citeauthor{traficante_massive_2018}
   \citeyear{traficante_massive_2018}, and the review by \citeauthor
   {motte_high-mass_2018} \citeyear{motte_high-mass_2018}). In contrast, \citet
   {padoan_supernova_2017} (and references therein) propose that 'turbulent
   fragmentation', where supersonic turbulence is predominantly driven by supernovae 
   explosions, determines the evolution and fragmentation of star formation regions.
   
   Disentangling scenarios such as these requires one to accurately characterise 
   the dynamics of star-forming regions from the core scale ($\sim 0.1$ pc) to
   the molecular complex scale (a few tens pc), and to identify the 
   structures and their origin in relation with the rest of the molecular complex.
   Molecular line observations are best suited to study the dynamics of molecular
   clouds, and, due to instrumental limitations, such studies often restrict their 
   analysis to the scales probed by a given instrument to the chosen target.
   However, it has become increasingly common that ISM studies combine instruments 
   with various angular resolutions \citep{csengeri_atlasgal_2016, liu_top-scope_2018}, 
   in particular with the advent of interferometric facilities \citep
   {henshaw_unveiling_2017, hacar_alma_2018}. 
   Another strategy consists in using very large high-resolution observational 
   programmes to access several scales simultaneously \citep{pety_anatomy_2017, 
   nakamura_wide-field_2017, sun_cloud-scale_2018}. However, multi-scale 
   studies remain scarce in the context of the early phases of star formation, 
   and we are far from having a representative sample of molecular complexes 
   where the nature and history of the most prominent structures would be understood.

   Another limitation to progress in the understanding of star formation could 
   result from the lack of diversity in the target selection. The Planck team 
   has taken advantage of the excellent sensitivity and all-sky coverage of the 
   {\it Planck} telescope between 30 and 857 GHz \citep{tauber_planck_2010} to 
   compile the first all-sky, unbiased catalogue of candidate star-forming 
   regions, the Planck Galactic cold clump catalogue \citep[PGCC, ][]
   {montier_all-sky_2010, planck_collaboration_planck_2011, 
   planck_collaboration_planck_2016}. Several hundred PGCCs have been followed-up
   with the {\it Herschel} observatory in the frame of the open time key programme
   Galactic cold cores \citep[GCC, ][]{juvela_galactic_2012}, with an unbiased 
   target selection strategy in terms of Galactic longitude and latitude, 
   distance and mass, and leading to a sample of great diversity \citep
   {montillaud_galactic_2015}. The PGCCs were further followed-up by  several
   projects. The 'SMT All-sky Mapping of PLanck Interstellar Nebulae in the Galaxy'
   (SAMPLING) survey \citep{wang_first_2018} is an ESO
   public survey of PGCCs in \CO\, and \tCO\,(J=2-1). The first data release
   contains 124 fields with an effective resolution of $36\arcsec$, a channel width of
   0.33 km\,s$^{-1}$ and an RMS noise of $T_{\rm mb}<0.2$ K. The TOP-SCOPE project 
   \citep{liu_top-scope_2018} combines the 'TRAO Observations of PGCCs' (TOP) 
   science key programme of the Taeduk Radio Astronomy Observatory (TRAO), a 
   survey of the J=1-0 transitions of \CO\, and \tCO\, towards $\sim 2000$ PGCCs, 
   and the 'SCUBA-2 Continuum Observations of Pre-protostellar Evolution' (SCOPE), 
   a large programme at the JCMT telescope targeting the 850\,$\mu$m continuum 
   of $\sim 1000$ PGCCs \citep{eden_scope:_2019}. 

   In Montillaud et al. (2019, hereafter Paper I), we reported the analysis of 
   the dense core population in the star-forming region G202.3+2.5, a complex
   filament at the edge of the Monoceros OB1 molecular complex and part of the
   GCC {\it Herschel} follow-up. We selected this target because of (i) its complex, 
   ramified morphology, suggestive of a complex dynamics, (ii) its significant 
   star formation activity as evaluated by \citet{montillaud_galactic_2015} and 
   (iii) its location near the well studied open cluster NGC 2264 whose impact 
   on the evolution of G202.3+2.5 needs to be investigated. In the present paper, 
   the second in this series, we report the analysis of the relationship between
   the dynamics and star formation activity in G202.3+2.5. To conduct this study, 
   we analyse the dust emission in the far-IR observed by {\it Herschel} and in
   the mid-IR by WISE on a 0.5 deg$^2$ area covering a physical length of $\sim
   10$ pc along the filament. This is combined with the molecular gas emission
   observed in the millimetre range with both the IRAM 30-m telescope, for high
   angular and spectral resolutions ($\sim 25\arcsec$ and 0.06 km\,s$^{-1}$) in
   a limited area, and the TRAO 14-m telescope, in a larger area but with lower
   resolutions. 

   With this dataset, we now make a second step in the multiscale analysis of
   this region, covering from the core scale ($\sim 0.1$ pc) to the filament
   scale ($\sim 10$ pc). We analyse the density and thermal structure of the
   cloud from its dust and gas emissions, and correlate them with the distribution
   of cores whose characteristics were derived from their IR spectral energy 
   distribution (SED) and molecular line emission in Paper I. The velocity field 
   is examined and separated in different components which are used to infer the
   evolution of the cloud in conjunction with its environment.

   The paper is structured as follows. The Mon OB 1 region is presented in 
   the next part of this introduction. In Section~\ref{sec:observations} we 
   present the observational data set used in this study. The methods are
   explained in Section~\ref{sec:method}. Section~\ref{sec:results} gathers
   the factual results on the cloud structure, the core characteristics and
   the velocity field. These results are interpreted in terms of the evolution
   of the cloud in Section~\ref{sec:discussion}. Section~\ref{sec:conclusion}
   summarises our study.

   \begin{figure*}
      \includegraphics[width=\textwidth]{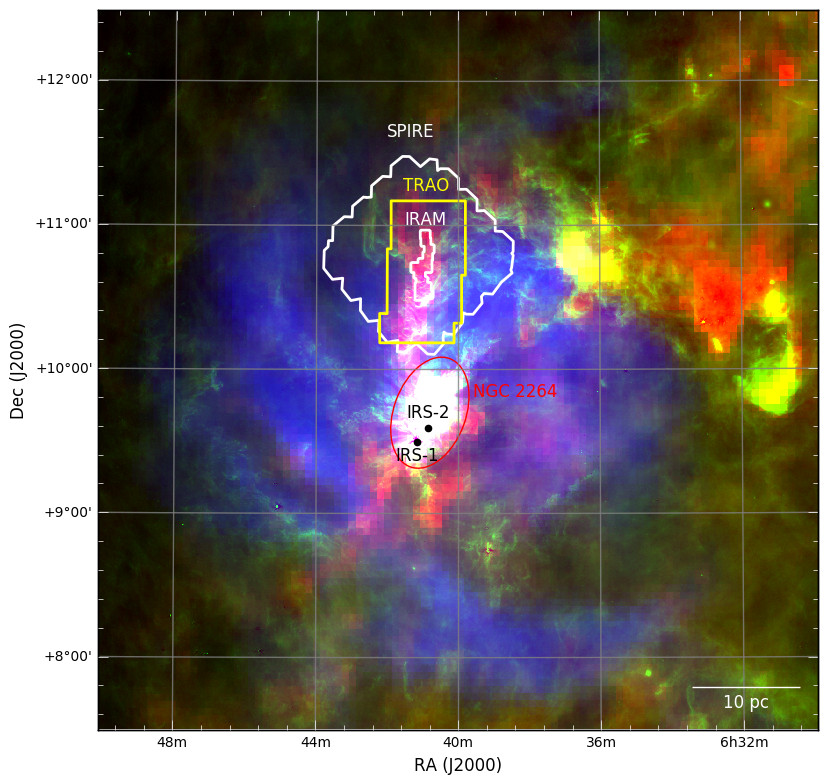}
      \caption{Three-colour composite image of Monoceros OB1 molecular complex.
      Red: {\it Planck} 857 GHz; Green: WISE 12\,$\mu$m from \citet{meisner_full-sky_2014};
      Blue: H$\alpha$. The white and yellow contours are the footprints of 
      {\it Herschel} SPIRE, IRAM, and TRAO observations, as labelled.
      }
      \label{fig:large_map}
   \end{figure*}

\section{The Monoceros OB1 molecular complex}

   The Monoceros OB1 molecular complex is a massive ($3.7 \times 10^4 \, M_{\odot}$)
   and relatively evolved ($>$3 Myr) star formation region \citep{dahm_young_2008}
   located 760 pc from the Sun in the approximate direction of the Galactic
   anti-centre. Thanks to this location, this sky region is almost devoid of 
   foreground and background emissions, with the exception of a few background
   clouds which belong to the Perseus Galactic arm more than 1 kpc farther, and
   can be identified from their radial velocities \citep[$\sim 20 - 40 \,
   \mathrm{km\,s^{-1}}$, ][]{oliver_new_1996}, significantly greater than those
   of the Monoceros OB1 molecular complex ($\sim 0 - 10 \,\mathrm{km\,s^{-1}}$).

   Next to the centre of the eastern part of this region lays the open cluster 
   NGC 2264 which contains more than 1000 members, the brightest of which is the
   O7 V multiple star S Monocerotis (S Mon). Along with several early B-type stars,
   they form the Mon OB1 association and are responsible for a vast H$\alpha$
   halo with $\sim 1.5$ degree in radius (20 pc) around NGC 2264 (blue area in
   Fig.~\ref{fig:large_map}). The H$\alpha$ survey by \citet{dahm_t_2005} revealed
   a population of several hundreds of T Tauri stars with ages scattered between
   0.1 and 5 Myr, indicating a sustained star formation activity. Numerous 
   embedded IR sources indicate that this activity is continuing. The first
   reported IR sources are IRS-1, or Allen's source \citep{allen_infrared_1972},
   a $9.5 M_{\odot}$ B2 zero-age main sequence star next to the Cone Nebula
   \citep{williams_gas_2002}, and IRS-2, an embedded cluster of protostars ~6'
   north of IRS-1 \citep{sargent_extended_1984, williams_gas_2002}. The active
   star formation within the region has been demonstrated by numerous
   identifications of outflows and Herbig-Haro objects from millimetre
   observations of molecular emission \citep{margulis_molecular_1988, 
   schreyer_molecular_1997, wolf-chase_star_2003} and narrow-band imaging of
   atomic lines \citep{ogura_giant_1995, reipurth_deep_2004}, as well as infall
   motions \citep{peretto_probing_2006}. More recently, several thousand young
   stellar object (YSO) candidates were identified from \textit{Spitzer} 
   observations of Mon OB1 east \citep{rapson_spitzer_2014}, and tens of starless
   and protostellar dense cores were identified in a 0.6 deg$^2$ field in the 
   northern part of the cloud from sub-millimetre \textit{Herschel} observations
   \citep{montillaud_galactic_2015}.

   In projection, NGC 2264 appears superposed on a wide (15'-30', corresponding
   to ~3-6 pc) elongated molecular cloud which extends over more than 2 degrees
   ($>25$ pc), as revealed, for example, by the sub-millimetre emission observed by the
   \textit{Planck} satellite at 857 GHz (red structure in Fig.~\ref{fig:large_map})
   or by the large scale \tCO\, map presented by \citet{reipurth_h_2004} (their
   Fig. 7). This latter map suggests a complex dynamics of the gas, but the
   low spatial and spectral resolutions (100\arcsec and 0.6 km\,s$^{-1}$, respectively) prevent
   one from a detailed analysis.

   \citet{rapson_spitzer_2014} showed that YSOs are distributed all over the 
   molecular complex, with most objects concentrated in NGC 2264, and the
   remainder following mostly the shape of the elongated molecular cloud. From
   the distributions of the various classes of YSOs in the cloud, they conclude
   that star formation in the Mon OB1 east molecular cloud is heterogeneous, with
   the star formation in the cloud being more recent than that in NGC 2264. 
   An even older, more dispersed population of stars may explain
   the large number of Class III objects in the region. Overall, this can be
   summarised as a gradient in star formation activity, which peaks in the 
   open-cluster NGC 2264 and systematically decreases towards the northern part
   of the complex.

   So far, most attention has been devoted to the brightest locations of the 
   region, namely the open cluster NGC 2264, IRS-1 and IRS-2, and the sources in
   their close surroundings. In the present paper, we focus on G202.3+2.5, the 
   northern tip of the Mon OB1 east molecular cloud, where \citet{montillaud_galactic_2015}
   report an active but recent star formation activity, and a complex morphology.

%=======================================================================================================================
% Observations
\section{Observations}\label{sec:observations}

\subsection{{\it Herschel} observations}\label{sec:obs_herschel}

   The cloud G202.3+2.5 was mapped with the instruments SPIRE
%   \footnote{
%   SPIRE has been developed by a consortium of institutes led by Cardiff 
%   University (UK) and including Univ. Lethbridge (Canada); NAOC (China); CEA, 
%   LAM (France); IFSI, Univ. Padua (Italy); IAC (Spain); Stockholm Observatory
%   (Sweden); Imperial College London, RAL, UCL-MSSL, UKATC, Univ. Sussex (UK);
%   and Caltech, JPL, NHSC, Univ. Colorado (USA). This development has been 
%   supported by national funding agencies: CSA (Canada); NAOC (China); CEA,
%   CNES, CNRS (France); ASI (Italy); MCINN (Spain); SNSB (Sweden); STFC (UK); 
%   and NASA (USA).}
   \citep[250\,$\mu$m, 350\,$\mu$m and 500\,$\mu$m, ]{griffin_herschel-spire_2010} 
   and PACS
%   \footnote{
%   PACS has been developed by a consortium of institutes led by MPE (Germany) 
%   and including UVIE (Austria); KU Leuven, CSL, IMEC (Belgium); CEA, LAM 
%   (France), MPIA (Germany), INAF-IFSI/OAA/OAP/OAT, LENS, SISSA (Italy); IAC 
%   (Spain). This development has been supported by the funding agencies BMVIT
%   (Austria), ESA-PRODEX (Belgium), CEA/CNES (France), DLR (Germany), ASI/INAF
%   (Italy), and CICYT/MCYT (Spain)} 
   \citep[100\,$\mu$m and 160\,$\mu$m, ]{poglitsch_photodetector_2010} onboard
   the \textit{Herschel} space observatory, as part of the \textit{Herschel} 
   open time key programme Galactic cold cores \citep{juvela_galactic_2010}.
   The characteristics and reduction steps for these maps are presented in detail
   in the GCC papers \citep[see for example ][]{juvela_galactic_2012}. The map
   resolutions are 18$\arcsec$, 25$\arcsec$ and 37$\arcsec$ for the 250\,$\mu$m, 
   350\,$\mu$m and 500\,$\mu$m bands
   of SPIRE, and of 7.7$\arcsec$ and 12$\arcsec$ for the 100\,$\mu$m and 
   160\,$\mu$m bands of PACS. The
   calibration accuracies of the {\it Herschel} SPIRE and PACS surface brightness
   are expected to be better than 7\%\footnote{SPIRE Observer's manual,\\
   http://herschel.esac.esa.int/Documentation.shtml} and 10\%\footnote{
   http://Herschel.esac.esa.int/twiki/bin/view/Public/PacsCalibrationWeb},
   respectively.

\subsection{TRAO observations}\label{sec:obs_trao}

   Large-scale maps of G202.3+2.5 were obtained in April 2017 with the 14-m 
   telescope of the TRAO as part of the Key Science Program named TOP (P.I. Tie 
   Liu). The SEQUOIA-TRAO, a multi-beam receiver with $4 \times 4$ pixels, was
   operated at 110.2 GHz with a spectral resolution of $\sim 0.04$ km\,s$^{-1}$
   and a beam size of 47\arcsec, to observe the \tCO\,(J=1-0) rotational transition.
   After smoothing the spectra to an effective resolution of $0.3$ km\,s$^{-1}$,
   the achieved sensitivity is $RMS(T_a^*) \approx 0.2 - 0.3$ K.

\subsection{IRAM observations}\label{sec:obs_iram}

   Part of the G202.3+2.5 cloud was observed with the IRAM 30-m telescope during 
   March 2017. We observed this region at the frequency of 110 GHz with the EMIR
   receiver to record the lines of \tCO\,(J=1-0) and \CeO\,(J=1-0). This front-end
   was connected to the VESPA autocorrelator configured to provide a spectral 
   resolution of 20 kHz, corresponding to 0.055 km\,s$^{-1}$ at 110 GHz. The FTS 
   autocorrelator was also connected in parallel with a spectral resolution of 
   200 kHz, enabling us to detect additional lines including the $^{12}$CO (J=1-0), 
   CS (J=2-1) and \NtHp (J=1-0) lines. Table~\ref{tab:lines} summarises the
   observations.

   We observed 14 tiles of typically $200\arcsec \times 180 \arcsec$ (i.e. 10 
   arcmin$^2$), and built a mosaic which covers some $130$ arcmin$^2$ around
   the position $(\alpha,\delta)_{J2000} = (6^h41^m00.5^s, +10^{\circ}42\arcmin
   27\arcsec)$. Each tile was observed multiple times and in orthogonal directions
   in on-the-fly (OTF) mode and position switching mode, with a scan velocity of
   at most 9\arcsec/s, a dump time of 1s and a maximum row spacing of 12\arcsec.
   The beam FWHM ranges from 21\arcsec at 115 GHz to 26\arcsec at 93 GHz.
   The off position was observed every 1 to 1.5 minutes. It was chosen at 
   $(\alpha,\delta)_{J2000} = (6^h42^m30.36^s, +10^{\circ}33\arcmin 21.2\arcsec)$, 
   after searching the SPIRE 250\,$\mu$m map for a minimum in surface brightness 
   (30.7 MJy/sr, to be compared to the values in the range 100-5000 MJy/sr in the 
   observed area). Pointing corrections and focus corrections were preformed
   every 1.5h and 3h, respectively, leading to a pointing accuracy measured to 
   be $\lesssim 5\arcsec$. We converted the antenna temperature to main beam 
   temperature assuming a standard telescope main beam efficiency\footnote
   {see http://www.iram.es/IRAMES/mainWiki/Iram30-mEfficiencies} of 0.78 for CO
   observations and 0.80 for CS and \NtHp.

   \begin{center}
   \begin{table}
      \caption{List of detected lines in our IRAM observations. }
      \begin{tabular}{c | c | c | c | c | c}
         \toprule
                           &  $\nu$    & Backend      & $\Delta v$      &  \Tmb$^{\rm max}$   & rms    \\
                           &   GHz     &              &  km s$^{-1}$    &    K                &   mK    \\
         \midrule
         \CO               &  115.271  & FTS          &  0.52           &  25.3          &  50 - 100    \\
        $\mathrm{C^{17}O}$ &  112.360  & FTS          &  0.53           &   1.2          &  50 - 100    \\
         \tCO              &  110.201  & FTS          &  0.54           &  10.8          &  50 - 100    \\
                           &           &     VESPA    &       0.054     &  11.9          &  150 - 300   \\
         \CeO              &  109.782  & FTS          &  0.54           &   2.6          &  50 - 100    \\
                           &           &     VESPA    &       0.054     &   4.1          &  150 - 300   \\
         CS                &  97.9809  & FTS          &  0.61           &   3.7          &  50 - 100    \\
        $\mathrm{C^{34}S}$ &  96.4129  & FTS          &  0.62           &   1.0          &  50 - 100    \\
         \NtHp             &  93.1763  & FTS          &  0.64           &   1.8          &  50 - 100    \\
         \bottomrule
      \end{tabular}
      \tablefoot{ The beam FWHM 
      ranges from 21\arcsec at 115 GHz to 26\arcsec at 93 GHz.
      The columns are:
      (1) Name of the species. All the transitions are J=1-0, except CS and 
      \CtS\, which are J=2-1. 
      (2) Frequency of the transition, from the CDMS database \citep{endres_cologne_2016},
      except for \NtHp, from \citet{pagani_frequency_2009}. For transitions with
      multiple components, $\nu$ is given for the component with the highest
      frequency.
      (3) Spectrometer used to record the data. FTS and VESPA were used at 
      resolutions of 200 kHz and 20 kHz, respectively.
      (4) Velocity resolution. All spectra were sampled with channels of 
      0.6 km\,s$^{-1}$ (FTS) or 0.06 km\,s$^{-1}$ (VESPA).
      (5) Maximum main beam temperature of the transition in the map. 
      (6) Approximate rms range computed from a 10 km\,s$^{-1}$ range, at least 20 km\,s$^{-1}$
      off the line, where no astronomical signal is found.
      }
      \label{tab:lines}
   \end{table}
   \end{center}

\subsection{Other observations}\label{sec:obs_other}

   The previous data sets are complemented with archival data. Large scale 
   H$_\alpha$ emission at 656 nm from the composite full-sky map by \citet
   {finkbeiner_full-sky_2003} is used in Fig.~\ref{fig:large_map} and in 
   Section~\ref{sec:discussion}. In the same section we also make use of the
   Second Digitized Sky Survey \citep[DSS, ][]{mclean_status_2000} in blue 
   ($\lambda=471$ nm), originating from the Palomar Observatory - Space Telescope 
   Science Institute Digital Sky Survey. The resolution is better than 2\arcsec,
   and data are given in scaled densities.

   In the mid-IR domain, we used data from the Wide-Field Infrared Survey 
   Explorer (WISE) satellite \citep{wright_wide-field_2010}. It has four bands 
   centred at 3.4, 4.6, 12, and 22\,$\mu$m with spatial resolution ranging 
   from 6.1\arcsec at the shortest wavelength to 12\arcsec at 22\,$\mu$m. We used
   these data to complement the SEDs of cores in the mid-IR range. The data were
   converted to surface brightness units with the conversion factors given in 
   the explanatory supplement \citep{cutri_explanatory_2011}. The calibration
   uncertainty is $\sim$6\% for the 22\,$\mu$m band and less for the shorter
   wavelengths.

%=======================================================================================================================
% Method
\section{Method}\label{sec:method}

%\subsection{Analysis of dust emission}

\subsection{Dust temperature and column density}

   The three SPIRE maps were combined, as in \citet{montillaud_galactic_2015},
   to compute maps of dust temperature \Tdust~ and column density \NHH, with an
   accuracy better than 1\,K in \Tdust~ in cold regions, corresponding to $\sim 20\%$
   in column density at 15\,K \citep{juvela_galactic_2012}. PACS data at shorter 
   wavelengths were not included in the calculation because they may bias the
   column density estimate due to the contribution of stochastically heated grains 
   \citep{shetty_effect_2009, shetty_effect_2009-1, malinen_accuracy_2011, 
   juvela_degeneracy_2013}. In G202.3+2.5, this seems particularly relevant since 
   \Tdust~ remains below 15 K except in the lowest column density regions outside
   the filaments, and at two very compact locations corresponding to young 
   stellar sources.

   To compute the \Tdust~ and \NHH\, maps, the SPIRE maps were convolved to a 
   resolution of $38.5\arcsec$, slightly greater than that of the 500\,$\mu$m 
   map and reprojected on the same grid. For each pixel, the spectral energy 
   distribution (SED) was fitted by a modified black-body function:
   \begin{equation}
      I_\nu \propto B_\nu (\Tdust) \nu^\beta
   \end{equation}
   where $B_\nu (\Tdust)$ is the Planck function at temperature \Tdust, and the
   spectral index $\beta$ was kept at the fixed value of 2.0. 

   The column density was then derived using the formula
   \begin{equation}
      \NHH = \frac{I_\nu}{B_\nu(\Tdust) \kappa_\nu \mu_{\rm H_2} \mH}
   \end{equation}
   where \mH~is the mass of the hydrogen atom, $\mu_{\rm H_2}=2.8$ is the mean 
   particle mass per hydrogen molecule, and $\kappa_\nu = 0.1 {\rm cm}^2/{\rm g}\,
   (\nu/1000{\rm GHz})^\beta$ is the dust opacity suitable for high density 
   environments \citep{beckwith_survey_1990}.

\begin{figure*}
      \includegraphics[height=10cm]{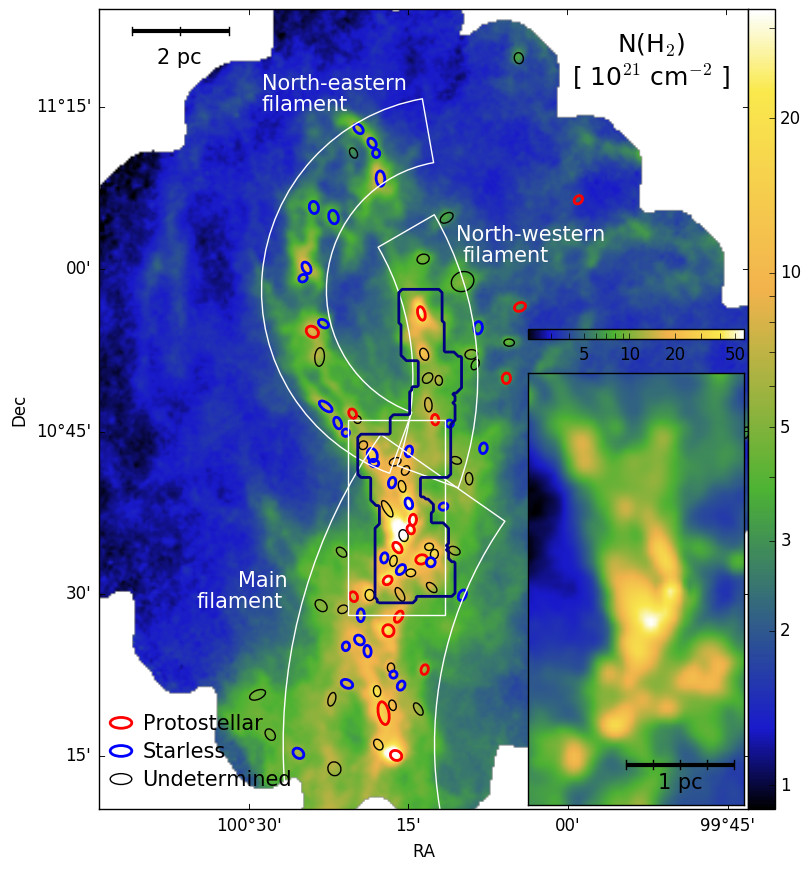}
      \hfill
      \includegraphics[height=10cm]{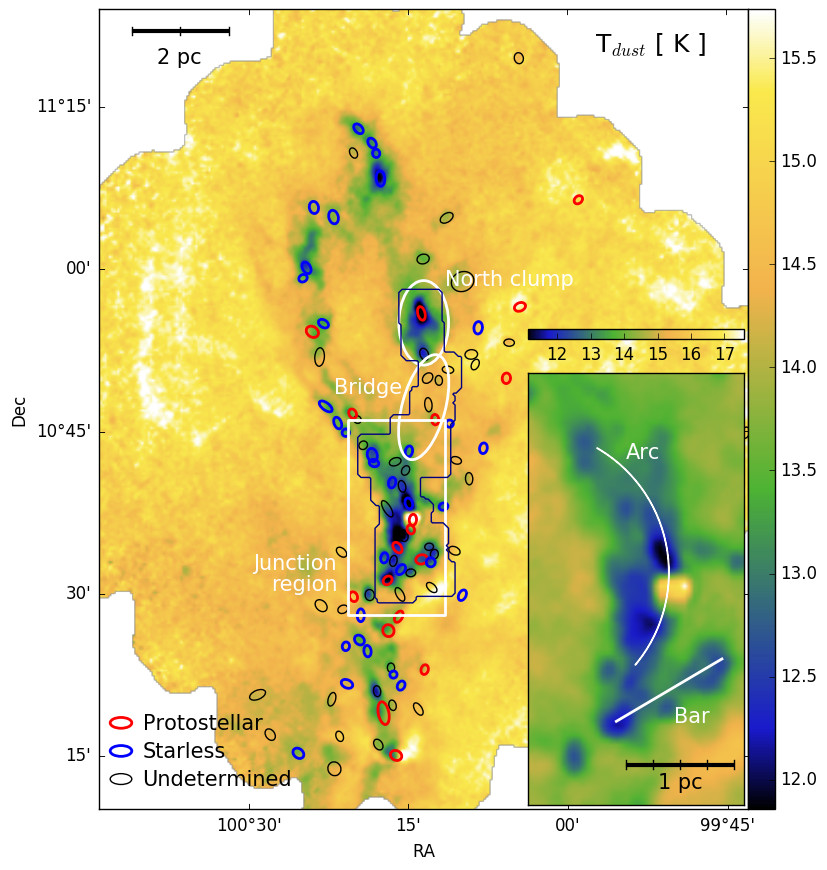}
      \caption{
      {\it Left:} Column density of molecular hydrogen in G202.3+2.5 as 
      derived from SPIRE bands, assuming a modified black-body SED with 
      fixed $\beta=2$.
      {\it Right:} Dust temperature in G202.3+2.5 derived from the same SED fitting. 
      In both frames, the dark-blue solid line shows the footprint of our IRAM OTF 
      observations, and the ellipses show the location, size (full-width at 
      half-maximum of a Gaussian fit) and orientation of the submillimetre
      compact sources extracted by \citet{montillaud_galactic_2015}. 
      The regions defined in the main text are indicated with white shapes.
      In both frames, the inset shows a zoom to the junction region (white
      rectangle).}
      \label{fig:map-Tdust-NH2}
   \end{figure*}

   The maps of \Tdust~ and \NHH~ obtained with this method are shown in 
   Fig.~\ref{fig:map-Tdust-NH2}.

\subsection{Analysis of the molecular line data}

   \subsubsection{Line fitting}
   \label{sec:linefitting}
   
   We computed pixel-by-pixel Gaussian fits of the entire IRAM maps. For
   each map independently, a first fit was done with a single Gaussian 
   function. If the residuals showed features with SNR$>$5, a new fit was tested
   with a two-component Gaussian function. Again, when the residual features had
   SNR$>$5, a three-component Gaussian fit was performed. This method is a 
   simplified version of the one adopted in Paper I for individual sources. The
   simplification was motivated by the large number of spectra to be fitted (several
   thousand spectra). As shown in Sect.~\ref{sec:res_velo_comp} and Fig.~\ref
   {fig:spec_13CO}, this is not sufficient for a few sightlines where up to four
   velocity components are seen. Still, we did not discard the few sightlines
   where a fourth component appears since for the vast majority of sightlines, 
   only one or two components dominate the spectra and were sufficiently well
   fitted (examples of the most complex spectra in this field are shown in 
   Fig.~\ref{fig:spec_13CO}) for the general analysis of velocity components.

   \subsubsection{Temperature and density calculations}
   \label{sec:TexColdens}

   We calculated the excitation temperatures \Tex~ and column densities of \tCO~
   and \CeO~ following the method described by \citet{wilson_tools_2013}. 
   We present our calculations in detail in the Sect.~A.2 of our Paper I. In short,
   we assume that the \CO\,(1-0) line is optically thick. It appears as a reasonable
   assumption since the H$_2$ column density is $\gtrsim 3 \times 10^{21}$ cm$^{-2}$ in 
   the area mapped with the IRAM 30-m (Fig.~\ref{fig:map-Tdust-NH2}). This enables
   us to compute directly, for each pixel, the excitation temperature of this 
   line from its maximum \Tmb. We then assumed that the J=1-0 transition of all
   the CO isotopologues have the same excitation temperature, so that we could
   derive the optical depths of \tCO\,(1-0) and \CeO\,(1-0), and therefore the
   column densities of \tCO\, and \CeO. The validity of this latter assumption is
   limited by the fact that the three transitions tend to probe different layers
   in the cloud. We present and discuss the column density maps in Sect.~\ref
   {sec:gas_coldens}.

\subsection{Velocity-coherent structures}

   We investigated the velocity-coherent structures in G202.3+2.5 as traced from
   the emission of CO isotopologues. We implemented a similar method as the FIVE
   algorithm developed by \citet{hacar_cores_2013}. Contrary to these authors,
   who analysed the emission of dense gas tracers (\CeO\, and \NtHp\, in \citet
   {hacar_cores_2013}, \NtHp\, and NH$_3$ in \citet{hacar_fibers_2017}), we focus
   on moderate density tracers (\tCO~ and \CeO) for the following reasons. 
   At 760 pc, G202.3+2.5 is farther than the clouds studied by Hacar and coworkers:
   238 pc for NGC 1333 \citep{hacar_fibers_2017} and 150 pc for Musca 
   \citep{hacar_musca_2016}. This implies that the dense structures are not
   as well resolved as in those studies. On the other hand, this greater distance
   enables us to cover a larger area, an asset to study the large scale dynamics
   in the cloud. CO isotopologues trace the appropriate densities to analyse the
   connections between the different large structures in the cloud.

   As the first step, we convolved each plane of the data cubes with a Gaussian
   kernel with a FWHM of 25\arcsec\, to improve the signal-to-noise ratio.
   For each pixel with a spectrum peaking at values with SNR$>6$, we used
   the central velocity of each Gaussian component of the fits presented 
   in Sect.~\ref{sec:linefitting} to build a cube of discrete points in the 
   position-position-velocity space, as shown in Fig.~\ref{fig:PPV-centroids}. 
   
   We used a friends-of-friends algorithm to connect the most related points and
   identify the coherent structures. For a given selected pixel, the neighbours
   in a box of $5 \times 5$ pixels and five channels were examined, corresponding 
   to a maximum radius of two cells around the selected pixel in each axis. The 
   box size, corresponding to 25\arcsec~ ($\sim 0.1$ pc), is similar to the beam
   size, securing the spatial coherence of the structure. Similarly, five 
   channels correspond to 0.3 km\,s$^{-1}$, which is typically the width of the
   most narrow lines in \tCO\, and \CeO, securing the velocity coherence of the 
   structure. The cells with a SNR>6 were considered friends of the selected
   pixel. The method was iterated by considering each new friend as the new
   selected pixel, until no new friend was found. We present the results in 
   Sect.~\ref{sec:res_velo_comp}.

%=======================================================================================================================
% Results
   \begin{figure*}
      \includegraphics[width=\textwidth]{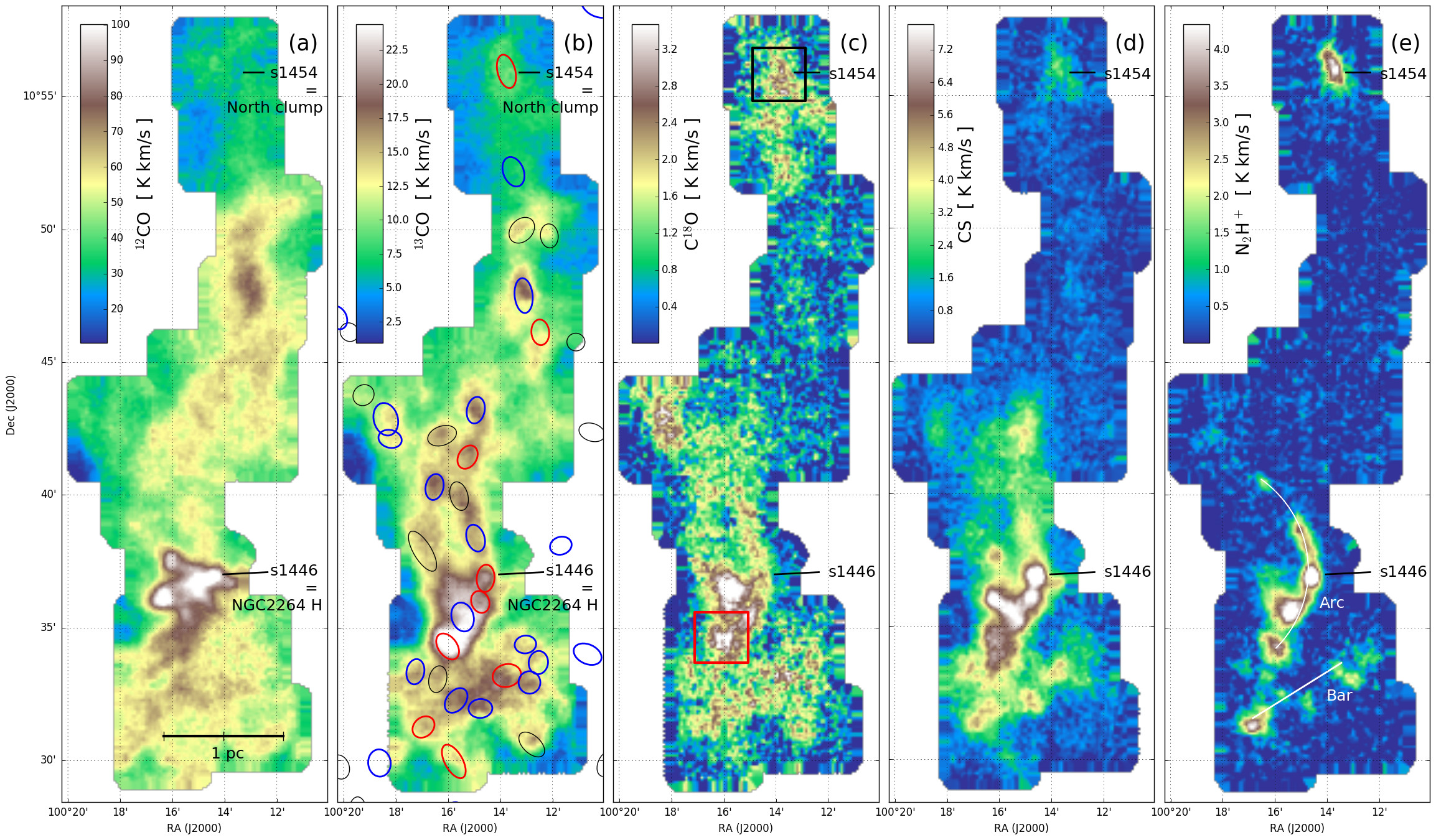}
      \caption{Integrated intensity maps of the main detected lines in our IRAM 
      observations. The sources s1454 (North clump) and s1446 (NGC 2264 H)
      are indicated in each frame. A scale bar is shown in frame (a). The ellipses
      in frame (b) show the GCC sources with the same colour code as in Fig.~\ref
      {fig:map-Tdust-NH2}: red for protostellar, blue for starless, and black
      for undetermined. The black and red squares in frame (c) show the areas
      where the average spectra in Fig.\ref{fig:COspectra_contrast} were computed.
      The structures named 'the arc' and 'the bar' are shown in frame (e).}
      \label{fig:integmaps}
   \end{figure*}

\section{Results}\label{sec:results}

   \begin{figure}
      \includegraphics[width=0.5\textwidth]{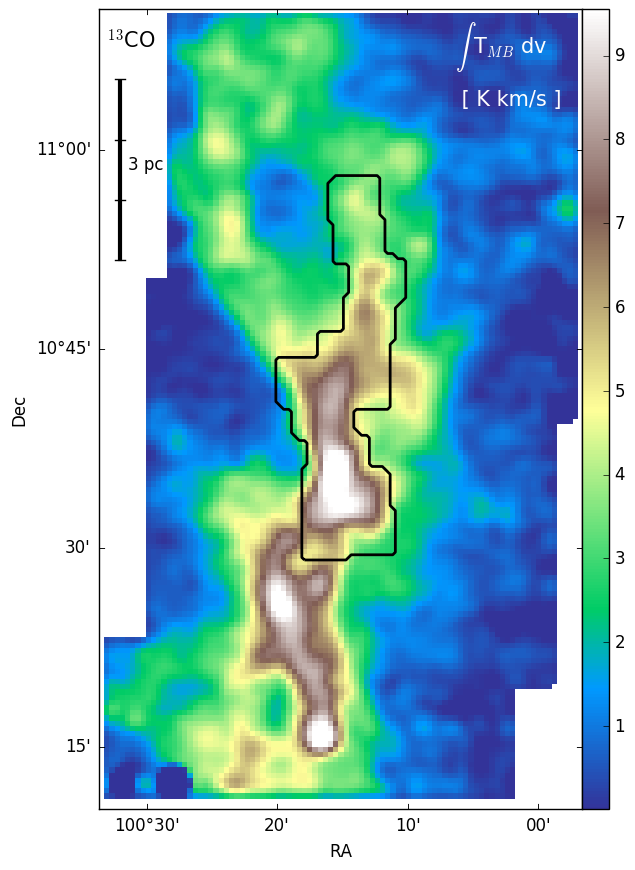}
      \caption{Integrated intensity map of \tCO\, in our TRAO observations.
      The black line shows the footprint of the IRAM observations. A scale
      bar is shown in the top left corner.}
      \label{fig:integmapTRAO}
   \end{figure}

   \begin{figure}
      \includegraphics[width=0.5\textwidth]{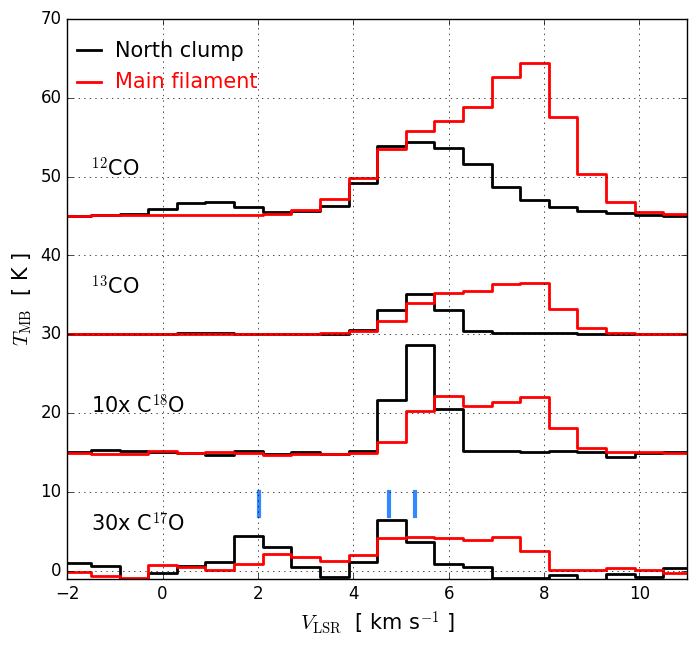}
      \caption{Spectra of CO isotopologues in the north clump (black) and in 
      the main filament (red). Spectra are averaged within 2 arcmin-wide square
      regions centred at ($\alpha$, $\delta$) = (6:40:54.9 ; +10:55:40.8) for 
      the north clump and ($\alpha$, $\delta$) = (6:41:03.6 ; +10:34:28.7) for 
      the main filament. In the \CsO\,(J=1-0) transition, the three vertical
      blue lines indicate the positions of the three hyperfine components
      assuming \Vlsr=5.3 km\,s$^{-1}$.}
      \label{fig:COspectra_contrast}
   \end{figure}

\subsection{The physical structure of G202.3+2.5}

   \subsubsection{Dust column density and temperature maps}
   \label{sec:res_dust_coldens}

   Figure~\ref{fig:map-Tdust-NH2} shows the distributions of the molecular 
   hydrogen column density \NHH\, and the dust temperature derived from \textit
   {Herschel} observations with a resolution of 38.5\arcsec, corresponding to 
   $\sim 0.15$ pc at the distance of 760 pc. It reveals the complex and ramified
   structure of G202.3+2.5, with two relatively broad ($\sim 0.5$ pc), cold 
   ($\Tdust < 14$ K), and dense ($\NHH \gtrsim 5 \times 10^{21} \, \mathrm{cm^{-2}}$) 
   filaments at $\delta > 10^\circ 45'$ (hereafter the north-western and 
   north-eastern filaments) joining into an even broader ($\sim 1$ pc) and 
   denser filamentary structure at $\delta < 10^\circ 45'$ (hereafter the main 
   filament; these structures are shown in the left frame of Fig.~\ref
   {fig:map-Tdust-NH2}). 
   
   The peak column density and lowest (line-of-sight average) temperature in the
   field are reached at the junction of the filaments with values $\NHH \approx 
   5 \times 10^{22} \, \mathrm{cm^{-2}}$ and $\Tdust \approx 11$ K. This junction
   region hosts a wealth of compact sources, many of which are aligned along an
   arc-shaped ridge (the arc in Fig.~\ref{fig:map-Tdust-NH2}), mostly 
   aligned with the north-south direction. This region corresponds to the 
   surroundings of the source labelled IRAS 27 or NGC 2264 H by \citet
   {wolf-chase_star_2003} where they found evidence of an outflow. 
   Interestingly, the peak \Tdust\, of the field is also reached in this area 
   with a value of $\Tdust \approx 17$ K, and is certainly an effect of
   the heating by the protostar responsible for the outflow. Just south of this
   source, a linear structure (the bar in Fig.~\ref{fig:map-Tdust-NH2}) inclined 
   from south-east to north-west joins the arc by its southernmost end, 
   contributing to the complexity of this ramified structure.

   A similar structure as that of NGC 2264 H is found at the southernmost end 
   of the main filament, although with less contrasted values of $\Tdust=13$\,K 
   and 17 K. This second source corresponds to the source labelled IRAS 25 or NGC 
   2264 O by \citet{wolf-chase_star_2003}, where they also found evidence of an
   outflow, and which \citet{ogura_giant_1995} identified as the object 
   responsible for the giant Herbig-Haro object HH 124 (outside of the \textit
   {Herschel} maps). These two objects are connected by the most central part, 
   or core in the sense defined by \citet{rivera-ingraham_galactic_2017}, of the
   main filament, where column density is $\gtrsim 10^{22} \, \mathrm{cm^{-2}}$ 
   and reaches values $\sim 2 \times 10^{22} \, \mathrm{cm^{-2}}$ and 
   temperatures $\Tdust \lesssim 13$ K at several compact locations. 

   The north-eastern filament exhibits a large arc-shaped and fragmented structure
   with column density generally $\lesssim 10^{22} \, \mathrm{cm^{-2}}$, except
   at a few compact locations where this value is marginally exceeded.

   The north-western filament is mostly composed of two regions: one large, dense,
   and cold clump (hereafter the north clump; top white ellipse in Fig.~\ref
   {fig:map-Tdust-NH2}, right) at its northernmost end ($\alpha$ = 
   6:40:54, $\delta$ = 10:55:52) with $\NHH \approx 2 \times 10^{22} \, \mathrm
   {cm^{-2}}$ and $\Tdust \approx 12.5$ K, and a more diffuse region which 
   connects the north clump to the junction region. Despite relatively large
   values of column density ($\NHH \sim 3-10 \times 10^{21} \, \mathrm{cm^{-2}}$), 
   this latter region (hereafter designated as 'the bridge', since it connects
   the north clump and the main filament, central white ellipse in Fig.~
   \ref{fig:map-Tdust-NH2}, right) has temperature values of $\sim$14.5 K, 
   comparable to those of the immediate surroundings of the cloud and making the
   north clump look isolated from the rest of the cloud in the temperature map
   of Fig.~\ref{fig:map-Tdust-NH2}.

   From these column density values, we derive the total masses and linear
   masses of the main structures discussed above as summarised in Table~\ref
   {tab:filament_masses}. We did not attempt any automatised extraction of the 
   filaments from the dust emission maps, since our molecular line observations 
   reveal a more complex structure than suggested by the dust emission. 
   We computed masses within column density contours of $2\times 10^{21}$ cm$^{-2}$ 
   and obtained values between 240 $M_\odot$ for the bridge and 2100 $M_\odot$ 
   for the southern part of the main filament. The junction region and the
   southern part of the main filament gather 3900 $M_\odot$, that is $\sim 60\%$
   of the cloud mass in the map. Table~\ref{tab:filament_masses} also lists 
   approximate estimates of the physical length of these structures, from which
   we derived linear masses between 120 and 230 $M_\odot$/pc in the northern 
   filaments, and of 530 and 580 $M_\odot$/pc in the junction region and southern
   main filament. These values are much larger than the critical linear mass for
   gravitational instability under thermal support. Using $M_{\rm lin,crit}=2 
   c_{\rm s}^2 /G$, where $c_{\rm s}$ is the isothermal sound speed and $G$ is 
   the gravitational constant \citep{ostriker_equilibrium_1964}, we obtain 
   $M_{\rm lin,crit} = \sim 16-32 \,M_\odot$/pc for a 10-20 K gas. However,
   turbulent and magnetic support can also contribute. We do not have constraints
   on the magnetic field strength in this paper, but in Paper I, we reported
   velocity dispersions $\sigma_{\CeO} \sim 0.4-0.6$ km\,s$^{-1}$ in the 
   junction region and 0.2-0.4 km\,s$^{-1}$ in the northern 
   filaments. Replacing the sound speed by $\sqrt{\sigma_{\rm NT}^2 + c_{\rm s}^2}$ in
   Ostriker's formula \citep[see Appendix~\ref{anx:turbsupport}]
   {chandrasekhar_fluctuations_1951} where $\sigma_{\rm NT}$ is the non-thermal 
   velocity dispersion (Eq.(A.6) in Paper I), we find critical linear masses of
   $M_{\rm lin, crit}= 92 - 186 \, M_\odot$/pc for the junction region and 
   $M_{\rm lin, crit}= 37 - 92 \, M_\odot$/pc for the northern filaments.

   This implies that, unless the magnetic field plays a major role, all the
   filaments should be fragmenting. This picture is consistent with the 
   numerous starless and protostellar cores identified everywhere along the 
   filaments. However the low threshold of $2\times 10^{21}$ cm$^{-2}$ includes 
   gas that is widely scattered and eventually may not, or not quickly, be 
   involved in star formation. A closer view of present or imminent star formation 
   is obtained by considering a greater column density threshold of $8\times 
   10^{21}$ cm$^{-2}$. This value corresponds approximately to an extinction
   of 8 mag, which is similar to the background extinction threshold of $\Av\sim 4-8$ 
   mag proposed by \citet{mckee_photoionization-regulated_1989} and reported by 
   \citet{enoch_comparing_2007} ($\Av\sim 8$ mag in Perseus) or \citet
   {andre_filamentary_2010} ($\Av\sim 10$ mag in Aquila). It 
   reveals that in the northern filament, only the north clump is clearly prone
   to star formation. With $M_{\rm lin}=130 M_\odot$/pc, the main filament is 
   also very active, but the junction region appears by far as the most active
   part of G202.3+2.5, with $M_{\rm lin}=210 M_\odot$/pc.

   \begin{table}
      \caption{Approximate lengths and masses of the main structures identified
      in the column density map.}
      \begin{tabular}{l c c c c c}
         \toprule
         Structure      & L  & $M_{\rm tot,2}$ & $M_{\rm tot,8}$ & $M_{\rm lin,2}$ & $M_{\rm lin,8}$ \\
         name           & [pc] & [$M_\odot$]   & [$M_\odot$]     & [$M_\odot$/pc]  & [$M_\odot$/pc]  \\
         \midrule
              North-eastern & 7.3 & 8.8(2) & 1.9(1) & 1.2(2) & 2.5(0) \\
              North-western & 4.2 & 9.6(2) & 7.7(1) & 2.3(2) & 1.8(1) \\
                North clump & 1.7 & 3.1(2) & 7.4(1) & 1.8(2) & 4.4(1) \\
                     Bridge & 1.5 & 2.4(2) & 2.8(0) & 1.6(2) & 1.9(0) \\
                   Junction & 3.5 & 1.8(3) & 7.3(2) & 5.3(2) & 2.1(2) \\
          Main filament$^a$ & 3.6 & 2.1(3) & 4.7(2) & 5.8(2) & 1.3(2) \\
         \bottomrule
      \end{tabular}
      \tablefoot{The columns are:
      (1) The name of the structure; 
      (2) The approximate length of the structure assuming a distance of 760 pc;
      (3) The total mass of the structure within a column density contour of 
      $2\times 10^{21}$ cm$^{-2}$;
      (4) The same for a $8\times 10^{21}$ cm$^{-2}$ \NHH\,contour;
      (5) The linear mass of the structure within a column density contour of 
      $2\times 10^{21}$ cm$^{-2}$;
      (6) The same for a $8\times 10^{21}$ cm$^{-2}$ \NHH\,contour;
      ($a$) Only the part of the main filament southwards of the IRAM footprint
      is accounted here. The properties of the complete main filament can be
      obtained by summing with the junction region. The numbers between parentheses
      are powers of 10.
      }
      \label{tab:filament_masses}
   \end{table}

   \subsubsection{Average spectra and integrated intensity maps}\label{sec:average_spectra}

   All the transitions summarised in Table \ref{tab:lines} are detected at least
   in the junction region. Figure~\ref{fig:integmaps} shows the velocity-integrated
   intensity maps of all the detected transitions in our IRAM dataset, except for C$^{17}$O 
   and \CtS\, which are only weakly detected in the junction region 
   ($\Tmb^{\mathrm{max}}\approx 0.3$ and 0.4 K, respectively) and in the north 
   clump ($\Tmb^{\mathrm{max}}\approx 0.4$ and 0.5 K, respectively, 
   corresponding to SNR $\sim 5$). The integrated maps of \CsO\, and \CtS\, 
   are shown in Appendix~A, Fig.~\ref{fig:integmaps_other}. A wider view of the
   \tCO\, emission in the region is presented in Fig.~\ref{fig:integmapTRAO}, 
   which shows the velocity-integrated map of the TRAO data.

   The \CO\, (J=1-0) emission (Fig.~\ref{fig:integmaps} a) fills the whole
   map with integrated intensities of the order of 50 K\,km\,s$^{-1}$ and exceeding 
   100 K\,km\,s$^{-1}$ around NGC 2264 H (s1446 in the GCC catalogue). It 
   presents a morphology mostly similar to the one observed from dust emission, 
   with the important exception of the north clump where significantly lower 
   values of the integrated intensities are found ($\sim 30$ K km\,s$^{-1}$). This 
   contrast in intensity between the north clump and the remainder of the field
   decreases when examining the emission of rarer CO isotopologues. Figure~\ref
   {fig:COspectra_contrast} compares the average spectra of the north clump and
   of a 2 arcmin square of the main filament (black and red squares, 
   respectively, in Fig.~\ref{fig:integmaps} c). It shows that the  ratio 
   of the peak \Tmb\, of CO isotopologues in the main filament, $\Tmb({\rm MF})$, 
   to that in the north clump, $\Tmb({\rm NC})$, decreases from $\Tmb({\rm MF})
   /\Tmb({\rm NC}) \sim 2$ for CO, to $\sim 1$
   for \tCO, $\sim 0.5$ for \CeO, and $\sim 0.8$ for \CsO. Moreover the line 
   width in the main filament is found to be about twice as large as that in the
   north clump for all isotopologues.

   The region of the bar (Fig.~\ref{fig:integmaps}.e) appears as a continuous 
   structure with a similar morphology for all the isotopologues (Fig.~\ref
   {fig:integmaps} a,b,c). In contrast, the arc region (Fig.~\ref{fig:integmaps}.e)
   shows different morphologies in all the CO maps, suggesting a complex structure 
   along the line-of-sight with variations in excitation possibly related to the
   outflow reported by \citet{wolf-chase_star_2003} and in Paper I.

   As shown in Fig.~\ref{fig:integmaps} d, the CS (J=2-1) emission peaks 
   at the location of NGC 2264 H with a peak value
   of 11.7 K\,km\,s$^{-1}$. The general morphology of the CS integrated intensity map is
   similar to that of the \CeO\, map, with (i) the junction region, and especially
   the arc which hosts NGC 2264 H and the bar, dominating the emission, (ii) the
   north clump showing relatively diffuse emission and (iii) the bridge not
   being detected. Differences are found in the arc, which appears more compact
   than in CO, and shows a bright extension to the north-east, and in the bar
   which appears fragmented in three clumps separated by $\sim 1.5-2.0$ arcmin
   ($\sim 0.35-0.45$ pc). The \CtS\, emission (Fig.~\ref{fig:integmaps_other})
   is about ten times fainter than that of CS but shows almost exactly the same
   distribution.

   The \NtHp\,(J=1-0) emission traces the densest regions of the cloud and
   therefore is seen at quite compact locations (Fig.~\ref{fig:integmaps} e). 
   The arc around NGC 2264 H is 
   very well traced by this transition with the map maximum value of 6.2 K\,km\,s$^{-1}$
   being reached at the location of NGC 2264 H. Another bright compact source
   appears in the same arc, 1.5 arcmin ($\sim 0.35$ pc) south of NGC 2264 H with
   a peak value of 5.0 K\,km\,s$^{-1}$. A 3-arcmin (0.7 pc) long, continuous portion of
   the arc is found with an intensity $\gtrsim 2.5$ K\,km\,s$^{-1}$. The bar, southwards
   of the arc, is fragmented into at least three compact sources, even more
   clearly than in the CS map. The north clump is bright in \NtHp\, with a peak
   integrated intensity of 4.0 K\,km\,s$^{-1}$, and a significant emission ($>1.0$ K\,km\,s$^{-1}$)
   within a $2 \times 1$ arcmin elliptical region.

   \subsubsection{Gas column density maps} \label{sec:gas_coldens}
\begin{figure*}
      \includegraphics[width=\textwidth]{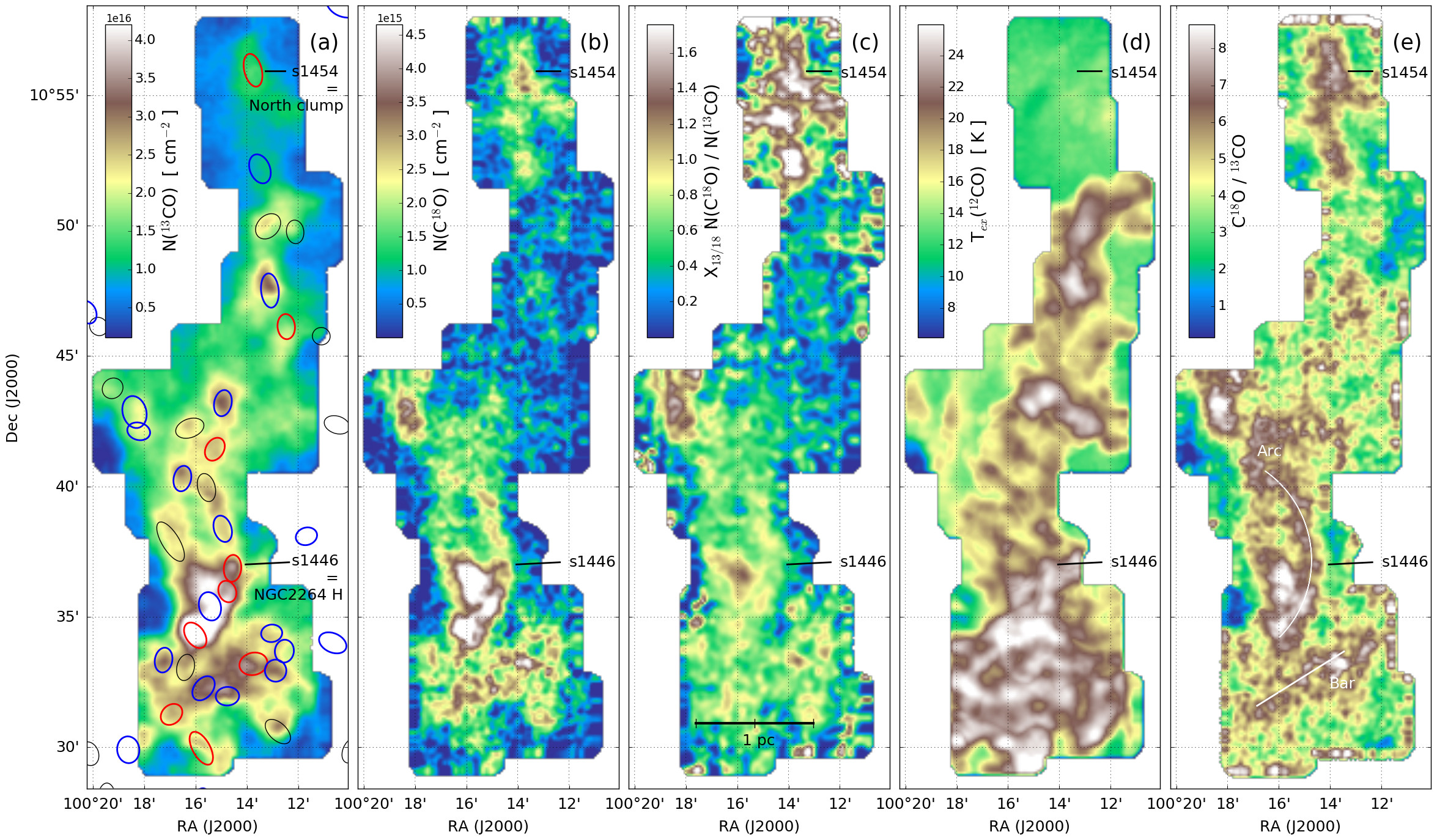}
      \caption{Column density maps of \tCO~ (a) and \CeO~ (b), and their ratio
      (c) normalised to the expected \tCO/\CeO\, from \citet{wilson_abundances_1994}.
      (d) Excitation temperature of \CO. (e) Integrated $\Tmb^{18}(\Vlsr)/\Tmb^{13}(\Vlsr)$
      ratio. The sources s1454 (north clump) and s1446 (NGC 2264 H) are 
      indicated in each frame. The ellipses in frame (a) show the GCC sources 
      with the same colour code as in Fig. 2: red for protostellar, blue for 
      starless, and black for undetermined. A scale bar is shown in frame (c).
      The structures named 'the arc' and 'the bar' are shown in frame (e).
      }
      \label{fig:Gcoldens}
   \end{figure*}

   Figure~\ref{fig:Gcoldens} compares the column density maps derived from IRAM
   observations as presented in Sect.~\ref{sec:TexColdens}. The same structures
   as seen in the dust column density map (Fig.~\ref{fig:map-Tdust-NH2}) are visible
   in the \tCO\, and \CeO\, column density maps, however with different contrasts.
   The values of $N(\tCO)$ (frame a) range between $5.0 \times 10^{15}\,\mathrm{cm^{-2}}$ 
   at the edge of the map and $5.3 \times 10^{16}\,\mathrm{cm^{-2}}$ at the peak
   of the junction region. To convert those values to H$_2$ column densities,
   we assume a ratio $X_{\mathrm{CO/H_2}}=1.0 \times 10^{-4}$ \citep[e.g. ][]
   {pineda_relation_2010} and we use the Galactic gradient in C isotopes 
   reported by \citet{wilson_abundances_1994} to compute the ratio $X_{\CO/\tCO}
   = 75.9$ for the galactocentric distance of G202.3+2.5 \citep[9.11 kpc, ][]
   {montillaud_galactic_2015}. We obtain \NHH\, values between $3.8 \times 10^{21}$ 
   and $4.0 \times 10^{22}\,\mathrm{cm^{-2}}$.

   In Fig.~\ref{fig:Gcoldens} b we also present the column density map for \CeO.
   The column density of $N(\CeO)$ reaches a maximum value of $6.6 \times 10^{15}\,
   \mathrm{cm^{-2}}$ in a different part of the junction region than \tCO. 
   Assuming a ratio $X_{\CO/\CeO}=618$ \citep{wilson_abundances_1994}, it 
   corresponds to a \NHH\, value of $4.1 \times 10^{22}$ cm$^{-2}$.

   The \NHH\, column density values obtained for \tCO\, and \CeO\, 
   are in good agreement with each other. This implies that \tCO\, emission is
   generally not sufficiently optically thick to strongly bias (underestimate)
   the \tCO\, column density measurement. Indeed, the peak values of $\tau_{\nu}$
   derived from Eq.(A.1) in Paper I for \tCO\, are $\sim 0.5$ in the 
   junction region, with maximum values of $\sim 1$ localised in compact 
   sources. Only the north clump exhibits larger values of $\sim 0.8$ with a
   maximum $\tau_{\rm peak} = 1.3$, despite column densities and peak \Tmb\, 
   being lower than in the southern area. 

   This is related to the trend discussed in Sect.~\ref{sec:average_spectra} and
   shown in Fig.~\ref{fig:COspectra_contrast}, where rarer isotopologues tend to
   be brighter in the north clump than in the junction region, while \CO\, is
   fainter in the north clump than in the rest of the filament. Similar trends
   are observed in Fig.~\ref{fig:Gcoldens} e, showing the integrated ratio of
   \CeO/\tCO\, cubes\footnote{To limit the noise in the map, only data cube cells
   with \Tmb\, greater than 1.5 K and 0.1 K in \tCO\, and \CeO, respectively, 
   were included. Using lower thresholds makes the map noisier, but does not 
   change the average values nor the morphology significantly.}, where the north
   clump has almost as high ratios as the junction region in spite of its
   fainter emission and lower column density. More striking is the map of 
   $N(\CeO)/N(\tCO)$ (Fig.~\ref{fig:Gcoldens} c), where the largest values are
   found in the north clump \citep[up to approximately two times the $X_{18/13}$ 
   value given by][]{wilson_abundances_1994}, and approximately three times 
   larger than in the rest of the 
   filament. Interestingly, the map of \CO\, excitation temperature (Fig.~\ref
   {fig:Gcoldens} d) reveals a strong contrast between the north clump 
   ($T_{\rm ex} \sim 12$ K, morphology practically invisible), and the rest of 
   the filament ($T_{\rm ex} \sim 15-30$ K, clear morphology which
   contrasts well with the background).

   The reasons for this large difference in \Tex\, between the north and south
   of the field are not fully clear. Apart from the north clump, the spatial
   variations in \Tex\, show a morphology very similar to the one of $N(\tCO)$.
   Since \Tex\, increases with increasing molecular hydrogen volume density \citep
   [see, for example, Fig.2.7 in ][]{yamamoto_introduction_2017}, and assuming 
   that the $N(\tCO)$ map is a good proxy for volume density variations, at 
   least part of the variations in \Tex\, are likely to reflect variations in 
   $n(\mathrm{H_2})$. Other possible effects contributing to these variations 
   include a lower external heating in the north clump and optical depth effects
   for example as a result of differences in velocity dispersion.

\subsection{Velocity components}
   \label{sec:res_velo_comp}

\begin{figure}
      \includegraphics[width=0.49\textwidth]{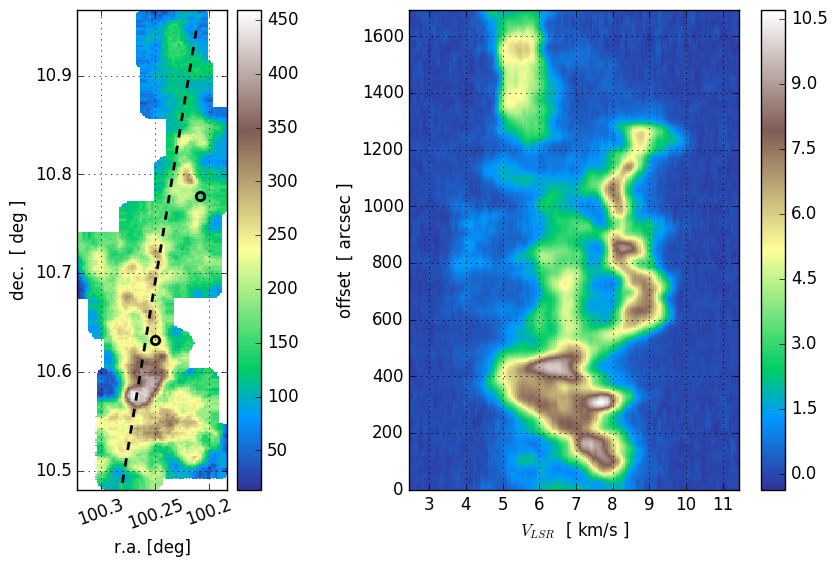}
      \caption{\textit{Left:} Velocity integrated map of \tCO\,in K\,km\,s$^{-1}$.
      \textit{Right:} Position-velocity diagram along the dashed line of the
      left frame. The two circles on the map in the left frame show the locations
      of the spectra shown in Fig.~\ref{fig:spec_13CO}.}
      \label{fig:PV-NS_13CO}
   \end{figure}

\begin{figure}
      \includegraphics[width=0.24\textwidth]{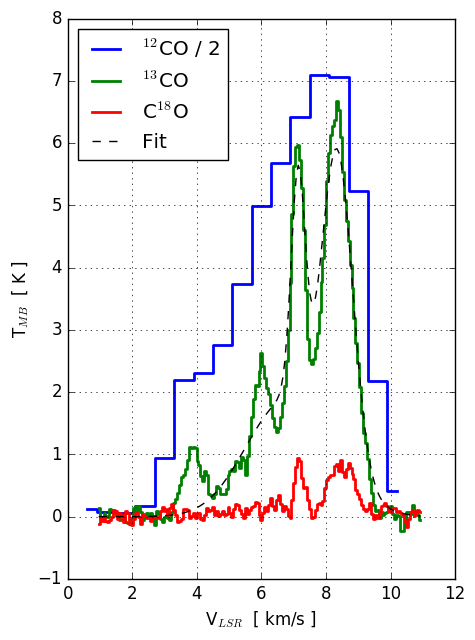}
      \includegraphics[width=0.24\textwidth]{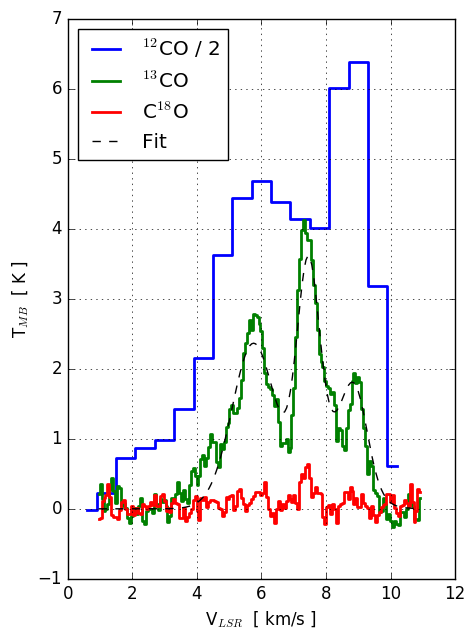}
      \caption{Spectra of CO, \tCO\, and \CeO\,(J=1-0) in the junction (left)
      and bridge (right) regions at the locations marked by black circles in 
      Fig.~\ref{fig:PV-NS_13CO}. The spectra of CO is divided by 2. The 
      black dashed lines are the 3-component Gaussian fit of the \tCO\, lines
      (Sect.~\ref{sec:linefitting}). }
      \label{fig:spec_13CO}
   \end{figure}

   The velocity structure of G202.3+2.5 is quite complex. Figure~\ref{fig:PV-NS_13CO}
   shows that in \tCO\, several velocity components are overlaid along most 
   sightlines, with a scatter in velocity greater than 4 km\,s$^{-1}$, and fluctuations
   in velocity within each component of $\sim 3$ km\,s$^{-1}$ in the junction region 
   (offset$<540\arcsec$), $\sim 3$ km\,s$^{-1}$ around the bridge (offsets between 540 and 
   1300\arcsec), and $<0.5$ km\,s$^{-1}$ in the north clump (offset$>$1300\arcsec). This complexity is
   even greater at some positions, as shown in Fig.~\ref{fig:spec_13CO}, where
   up to four different velocity components are found. 
   
\begin{figure*}
      \includegraphics[width=\textwidth]{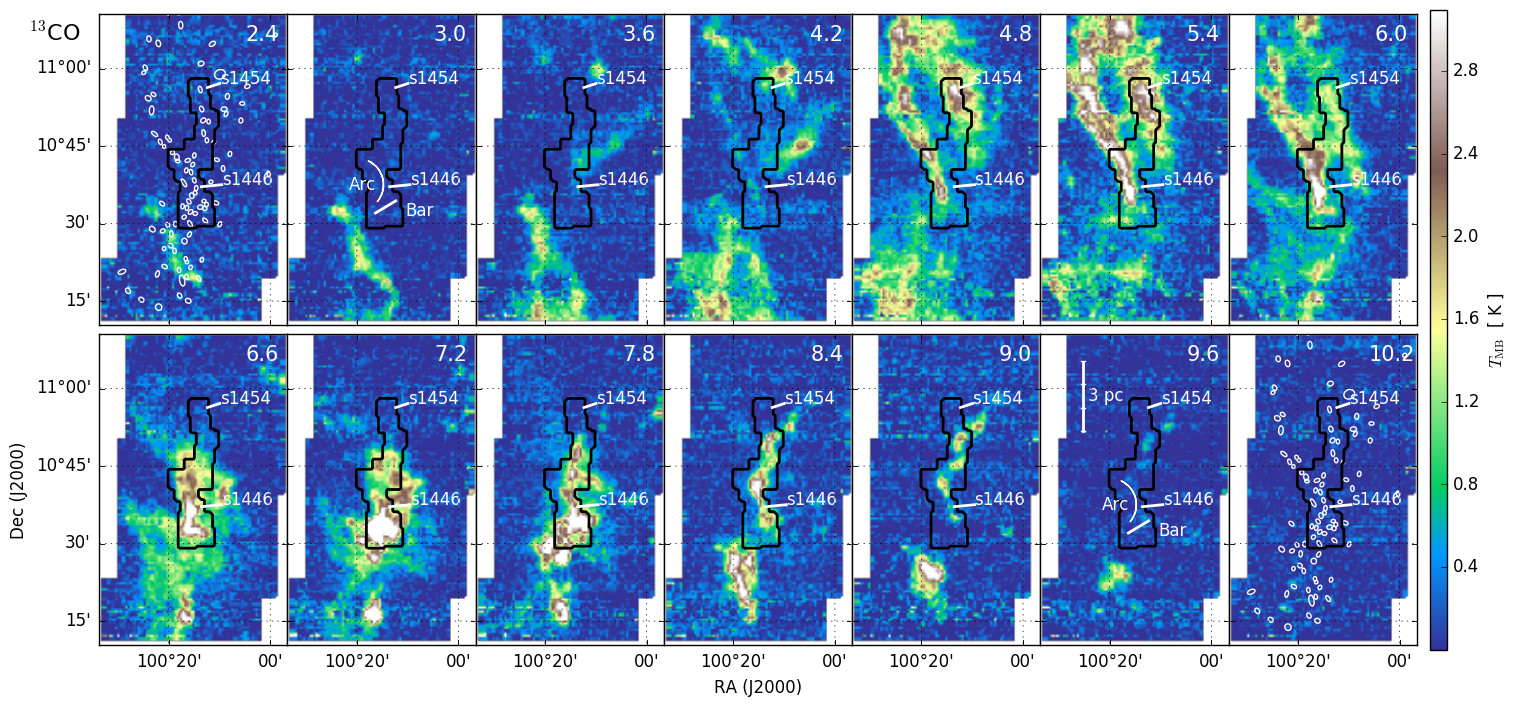}\\
      \includegraphics[width=\textwidth]{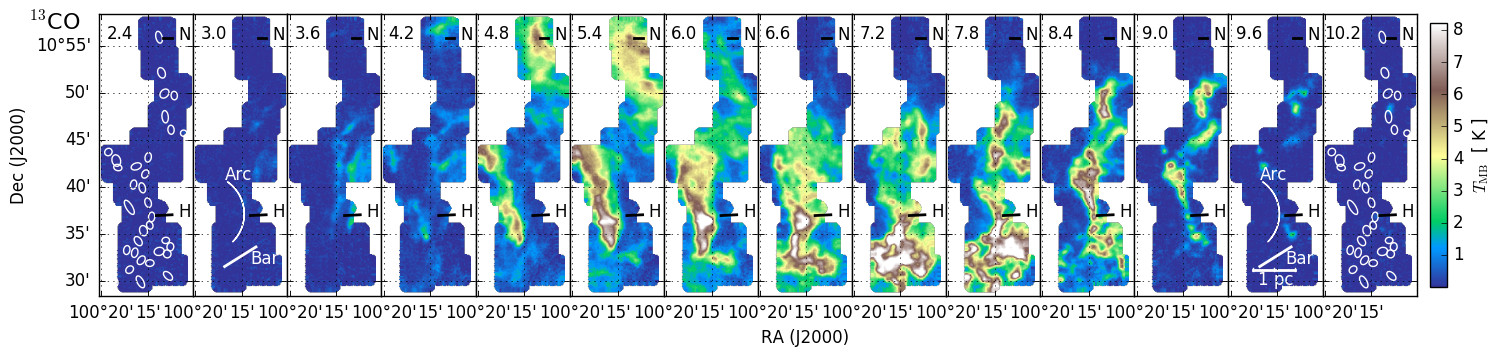}\\
      \includegraphics[width=\textwidth]{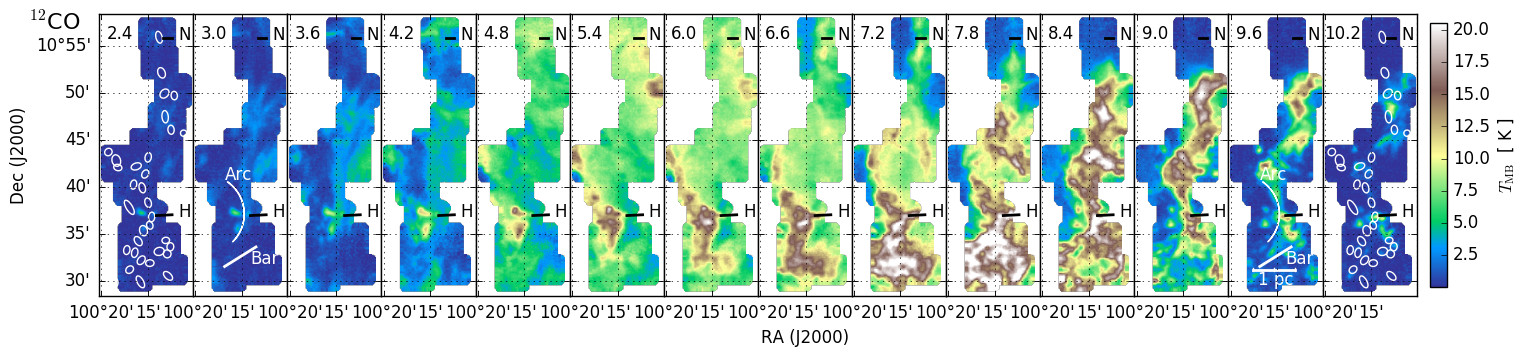}
      \caption{ 
      \textit{Top:}     \tCO\, channel maps from TRAO data.
      \textit{Middle:}  \tCO\, channel maps from IRAM data.
      \textit{Bottom:}  \CO\,  channel maps from IRAM data.
      The channel velocities are written in the top corners. In the two 
      first and last frames, the sources from the GCC catalogue are shown as 
      white ellipses and the arc and bar regions are shown as white lines. The 
      north clump and NGC 2264 H are labelled as 's1454' and 's1446' in TRAO 
      maps, and 'N' and 'H' in IRAM maps, respectively. Scale bars are in the 
      next to last frames.}
      \label{fig:channelmaps}
   \end{figure*}

   Figure~\ref{fig:channelmaps} shows the \tCO\, channel maps of G202.3+2.5 from
   TRAO and IRAM data, as well as the \CO\, channel maps from IRAM data. TRAO
   data reveal that the north-eastern and north-western filaments (including the north clump)
   are parts of the same large structure which appears as an elongated loop at
   velocities between 4 and 6 km\,s$^{-1}$. Both TRAO and IRAM data show that 
   between 6 and 8 km\,s$^{-1}$ the emission of the filament is dominated
   by the junction region and the main filament, and that at 6 km\,s$^{-1}$ the 
   brightest part of this structure is in the junction region, confirming that
   there is a continuity from the north-eastern and north-western filaments to the main
   filament. At larger velocities ($\Vlsr\gtrsim 7.5$ km\,s$^{-1}$), the
   main filament continuously extends to the north west, up to the region of the
   bridge. Interestingly, although in \tCO\, the north clump seems brighter than
   the bridge, in the IRAM \CO\, channel maps the situation is reversed. 
%   The latter
%   maps also show two ring-shaped structures, one in the junction region at
%   $\sim 7-8$ km\,s$^{-1}$, the second one in the southernmost part of the IRAM
%   map in the 8.4 km\,s$^{-1}$ channel. We discuss these structures in 
%   Sect.~\ref{sec:rings}. 
   The channel maps of the other tracers detected with 
   the IRAM telescope are shown in appendix~\ref{anx:channelmaps}.


\begin{figure*}
      \includegraphics[width=0.49\textwidth]{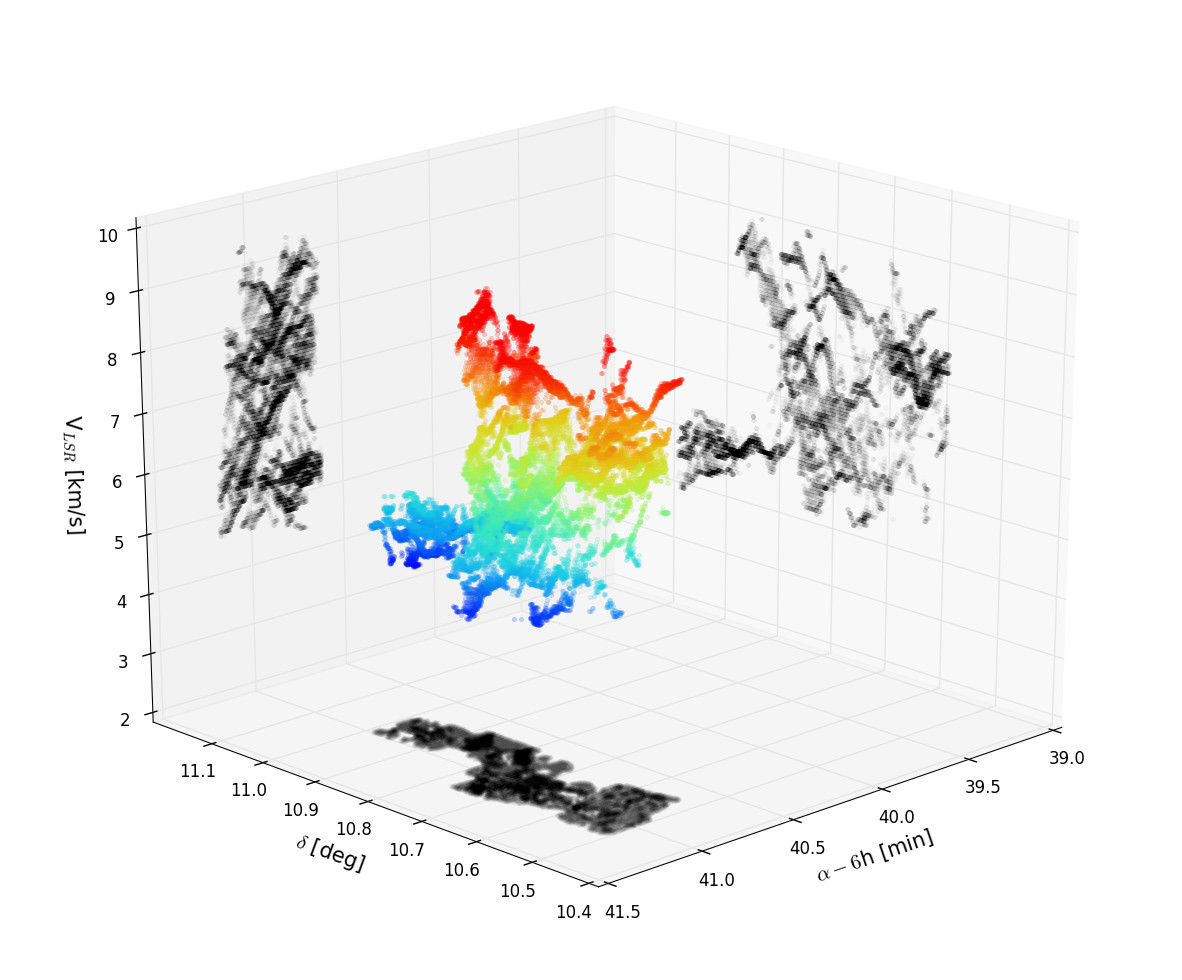}
      \hfill
      \includegraphics[width=0.49\textwidth]{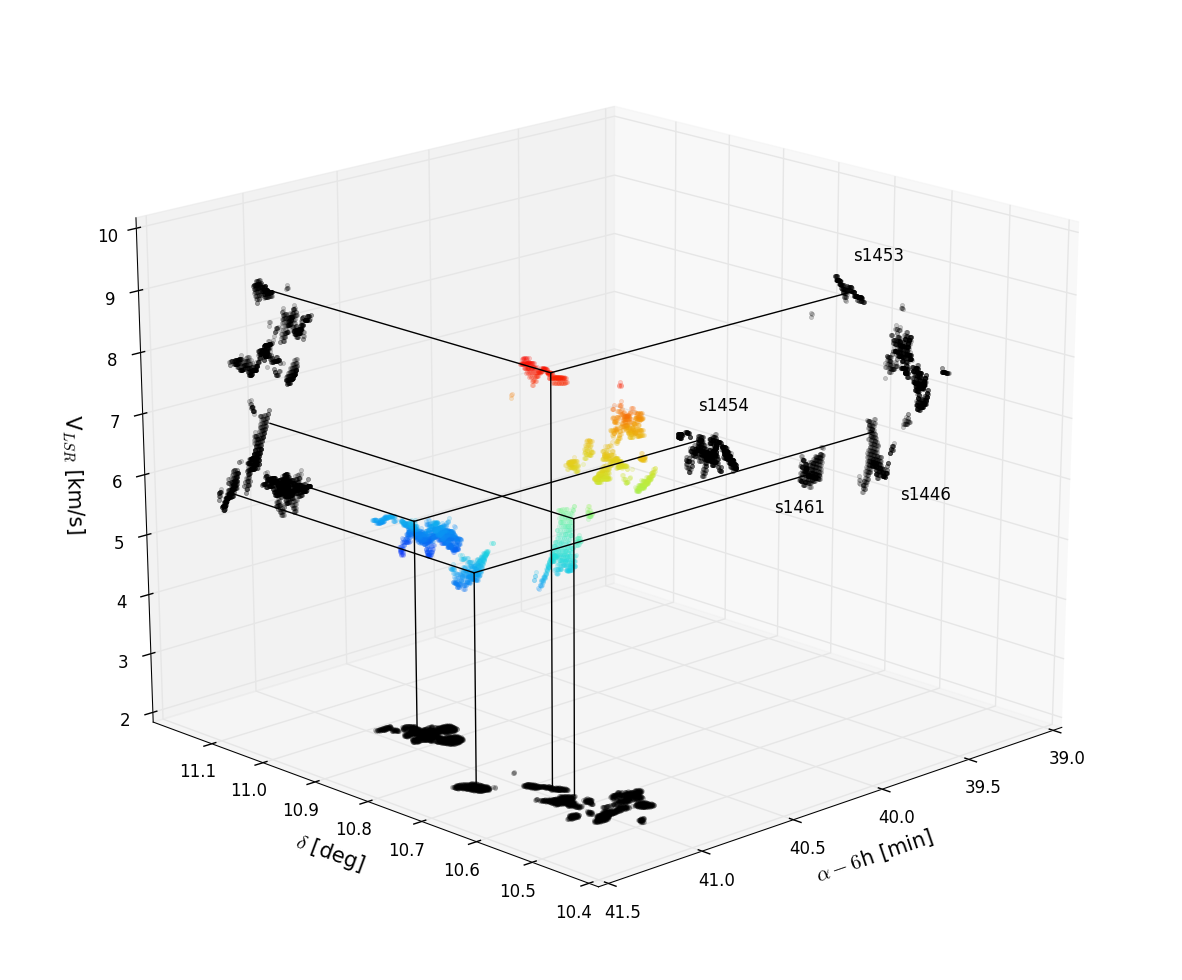}
      \caption{Distribution of line centroids in the position-position-velocity
      space for \tCO~({\it left}) and \CeO~({\it right}). The colours vary
      with velocity to help the 3D visualisation. The black points are projections 
      of the colour ones on the faces of the box. In the right frame, lines
      joining the projections to the colour points are drawn to help 
      visualising in 3D the positions of s1454 (in the
      north clump), s1461 (at the root of the north-eastern filament), s1446 (in
      the arc), and s1453 (just north to the arc).}
      \label{fig:PPV-centroids}
   \end{figure*}

   Another view of the velocity structures is presented in Fig.~\ref{fig:PPV-centroids}
   which shows the line centroids obtained from our \tCO\, and \CeO\, cubes. 
   The \tCO\, frame reveals multiple interconnected components and large-scale
   velocity gradients. The lower point density in the \CeO\, centroid distribution
   unveils the skeleton of the cloud, with a clear large-scale velocity gradient
   in the north-south direction, and a more confused east-west gradient
   in the junction region.
   
\begin{figure}
      \includegraphics[width=0.49\textwidth]{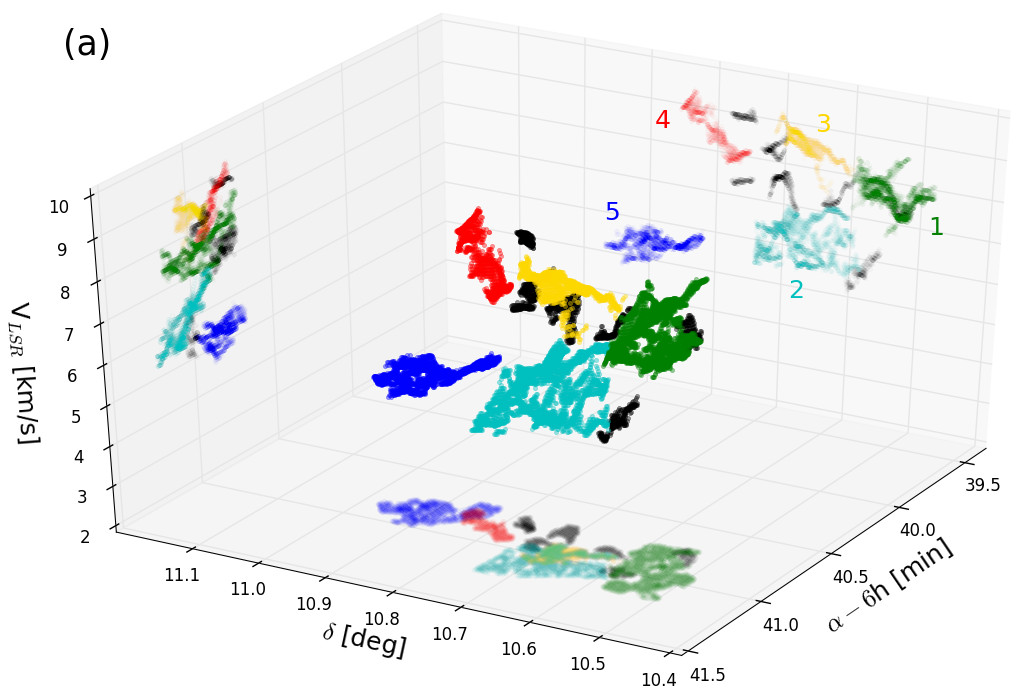}
      \hfill
      \includegraphics[width=0.49\textwidth]{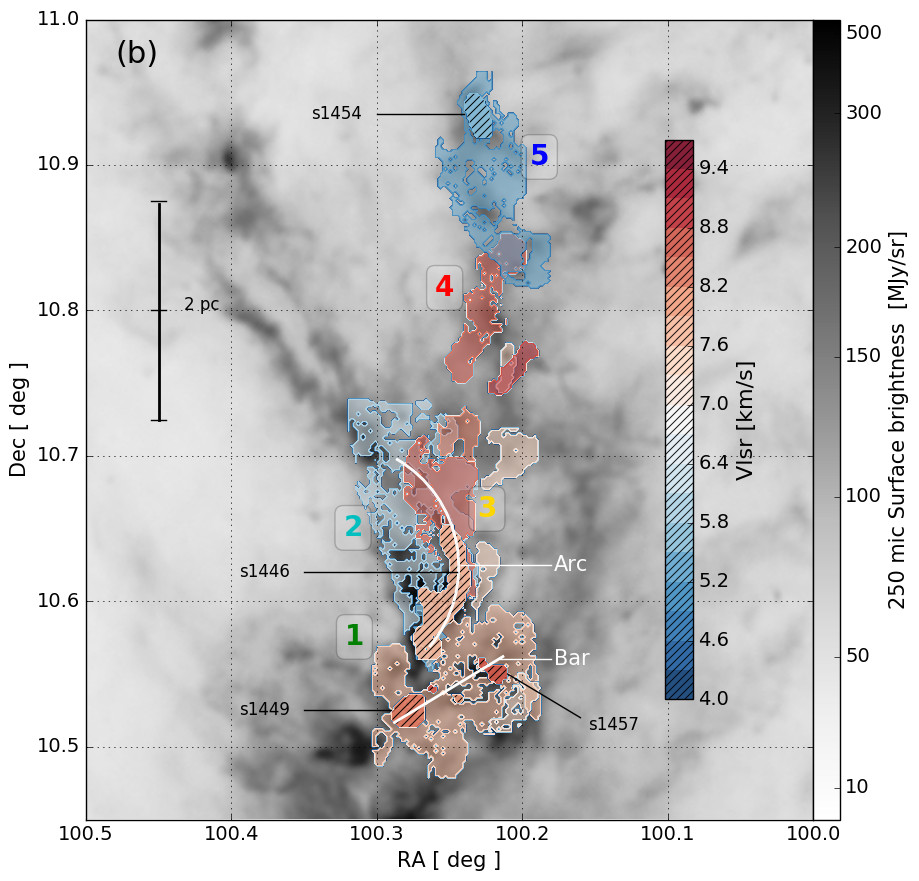}
      \caption{(a) 3D position-position-velocity view of the 
      velocity-coherent structures in $\mathrm{^{13}CO}$. The pale colours are 
      projections of the structures onto the plane of the sky, the $\alpha - v$
      plane and the $\delta -v$ plane. Each colour corresponds to a different 
      structure. The id numbers of the five components tabulated in Table~\ref
      {tab:velocomp} are indicated in the $\delta -v$ plane.
      (b) Thermal emission of dust in G202.3+2.5 at $\mathrm{250\mu m}$ 
      ({\it Herschel}/SPIRE). Coloured regions show the same velocity-coherent 
      structures as in frame (a) for $^{13}$CO J=1-0 (no symbol) and 
      \NtHp\, J=1-0 (isolated component, with hatches). The colour scale
      shows the average radial velocity of each structure. Transparency reveals
      the overlaps between the structures. Sources 1446, 1449, 1454, and 1457 
      are indicated in black. The white lines show the arc and the bar. The
      coloured numbers indicate the five main VCSs.}
      \label{fig:res_velo_comp}
   \end{figure}

   Figure~\ref{fig:res_velo_comp} shows the result of the friends-of-friends
   identification of \tCO\, velocity-coherent structures (VCS). Five major 
   structures are identified, three in the junction region (VCS 1-3), the two 
   others corresponding to the north clump (VCS 5) and the bridge (VCS 4, between
   the north clump and the junction region). The north clump appears as a single 
   structure at $\sim 5$ km\,s$^{-1}$
   relatively unattached to the junction region. The bridge, which seems to 
   connect the north clump to the junction region in {\it Herschel} maps, turns out
   to be completely disconnected in velocity from the north clump, and presents
   velocities similar to the reddest ones in the junction region, as well as a 
   significant velocity gradient of $\sim 0.3\, \mathrm{km\,s^{-1}\,arcmin^{-1}}$ (=1.5 
   km\,s$^{-1}$\,pc$^{-1}$). The yellow structure (VCS 3) in the PPV diagram of 
   Fig.~\ref{fig:res_velo_comp} a has similar velocities as the bridge and its
   morphology in the plane of the sky suggests that it could be its continuation 
   southwards. It strongly overlaps (in the plane of the sky) with the light 
   blue structure (VCS 2) in Fig.~\ref{fig:res_velo_comp} a, which spans bluer 
   velocities, from 5 km\,s$^{-1}$ (as in the north clump), to $\sim 7$ km\,s$^{-1}$ (the average 
   velocity of the cloud). These two structures are well separated by a gap in
   velocity which oscillates along the north-south direction, as seen in the 
   $\delta -v$ plane of Fig.~\ref{fig:PPV-centroids} for \tCO, and are mostly
   connected via VCS 1 at the southernmost end of the IRAM map. This is 
   particularly striking in the $\delta -v$ plane of Fig.~\ref{fig:PPV-centroids} 
   for \CeO.

   The second panel of Fig.~\ref{fig:res_velo_comp} shows the same structures 
   projected on the plane of the sky, with the average velocity of each VCS 
   colour coded. This figure also shows the densest parts of the cloud as
   traced by the \NtHp\, emission (hatched regions), for which we obtained the 
   velocity of the isolated component using hyperfine structure line fitting.
   The dense part of the arc appears to be located along the edge of VCS 2 (the 
   light-blue component), and next to VCS 1 (the light-red component). 
   With average velocities of about 7.7, 7.3, and 6.3 km\,s$^{-1}$ for VCS 1, the \NtHp\,
   emission of the arc, and VCS 2, respectively, the arc is also between VCS 1
   and 2 along the velocity axis. The other notable \NtHp\, emissions
   correspond to s1449 (\Vlsr = 7.56 km\,s$^{-1}$), s1457 (\Vlsr=7.81 km\,s$^{-1}$), and s1454
   (\Vlsr=5.29 km\,s$^{-1}$). Those sources are located in VCS 1 (\Vlsr=7.7 km\,s$^{-1}$), for 
   s1449 and s1457, and VCS 5 (\Vlsr=5.5 km\,s$^{-1}$), for s1454, both spatially and in
   the velocity space.

   Table~\ref{tab:velocomp} gives the main properties of the five largest \tCO\,
   velocity-coherent structures. Consistently with the results reported in Sect.~
   \ref{sec:gas_coldens} and Fig.~\ref{fig:Gcoldens}, the north clump (VCS 5) 
   stands out with a low average \tCO\, column density, a low scatter in \tCO\, 
   brightness temperature and a low radial-velocity dispersion. This suggests 
   that it belongs to a part of G202.3+2.5 with different environmental conditions
   than in the four other components. VCS 1-3 are all immediately next to the 
   junction region and around the dense gas traced by \NtHp. They individually 
   present large radial velocity ranges between $\Delta \Vlsr \sim$ 1.7 and 1.9
   km\,s$^{-1}$ with large \tCO\, column densities ($\gtrsim 7 \times 10^{21}$ cm$^{-2}$).
   Finally, the impression given by Fig.~\ref{fig:res_velo_comp} that VCS 4 (the 
   bridge) is the northwards continuation of VCS 3 is supported by the similarity 
   of their velocity dispersions, average and dispersion \Tmb(\tCO) and average 
   \tCO\,column densities.

   \begin{center}
   \begin{table*}
      \caption{Properties of the five largest \tCO\,velocity-coherent structures identified in G202.3+2.5.}
      \begin{tabular}{c c c c c c c c c c c c c }
      \toprule
      id & RA    & DEC    & $\langle \Vlsr \rangle$   & Area       & $\Delta \theta_{\rm min}$ & $\Delta \theta_{\rm max}$ & $e$ & $\Delta \Vlsr$ & $\langle T_{\rm mb} \rangle $ & $\sigma(T_{\rm mb})$ & $\langle N\mathrm{(H_2)} \rangle$ & $M$\\
        & (h m s) & (d m s)  & (km s$^{-1}$)          & (arcmin$^2$) & ($\arcmin$)                & ($\arcmin$)                  &     & (km s$^{-1}$)  & (K)                           & (K) & (cm$^{-2}$) & (M$_\odot$)\\
      \midrule
      1 & 06 40 59.7 & 10 32 36.5 &     7.77 & 25.72 &  3.54 &  8.75 &  2.47 &     1.92 &  7.76 &  1.84 &     10.5(21) &   292\\
      2 & 06 41 05.9 & 10 40 03.6 &     6.27 & 23.58 &  3.13 &  9.75 &  3.12 &     1.68 &  5.52 &  1.19 &      7.9(21) &   203\\
      3 & 06 41 01.0 & 10 40 02.4 &     8.63 & 10.26 &  1.96 &  6.96 &  3.55 &     1.80 &  6.74 &  1.29 &      7.5(21) &    84\\
      4 & 06 40 54.6 & 10 48 11.3 &     8.55 &  7.14 &  1.21 &  6.54 &  5.41 &     1.62 &  6.30 &  1.33 &      5.5(21) &    43\\
      5 & 06 40 53.1 & 10 53 25.4 &     5.51 & 17.95 &  3.21 &  8.83 &  2.75 &     1.20 &  4.77 &  0.81 &      5.2(21) &   103\\
      \bottomrule
      \end{tabular}
      \tablefoot{ The columns are, for each component:
      (1) The identity index;
      (2-3) The central coordinates;
      (4) The average radial velocity;
      (5) The total area in the plane of the sky;
      (6-7) The minimum and maximum extent;
      (8) The elongation, computed as the ratio of the two previous columns;
      (9) The velocity extent, computed as the difference between the minimum and maximum \Vlsr\, of the component;
      (10) The mean \tCO~main-beam temperature;
      (11) The standard deviation of \tCO~brightness temperature;
      (12) The mean column density from \tCO\,emission, where $(21)$ means $10^{21}$;
      (13) The mass from \tCO\,emission.
      }
      \label{tab:velocomp}
   \end{table*}
   \end{center}

%=======================================================================================================================
\section{Discussion}\label{sec:discussion}

   The goal of the paper is to investigate whether the dynamics of G202.3+2.5 
   shapes star formation in this cloud. In this section, we first investigate
   the outflows from the junction region. We then make an 
   inventory of star formation activity in the cloud, and then show that the
   cloud is made of two colliding filaments. We end by discussing the role of
   triggering in this star formation region.

%\subsection{Star formation activity}

%%\subsubsection{Rings}
%%   \label{sec:rings}

%%   In Sect.~\ref{sec:res_velo_comp} we mentioned two rings visible in the \CO\,
%%   channel maps. The first one is located in the junction region at
%%   $\sim 7-8$ km\,s$^{-1}$, the second one in the southernmost part of the IRAM
%%   map in the 8.4 km\,s$^{-1}$ channel. We searched for a central source next to
%%   the centre of each ring. We only found one for the

\subsection{Outflows from the junction region}
   \label{sec:outflow}

   Outflows from NGC 2264 H (source 1446) have been detected by \citet{
   wolf-chase_star_2003} from \CO(2-1) observations, with a complex morphology
   which is not clearly bipolar (see their Fig.~9). \citet{reipurth_deep_2004}
   showed from H$_{\alpha}$ images that the Herbig-Haro objects HH 576 (on the 
   north-west of NGC 2264 H) and HH 577 (on the south-west of NGC 2264 H) are
   at the end of flows that cross roughly at the location of this source (see
   their Fig.~7).

\begin{figure*}
      \includegraphics[width=\textwidth]{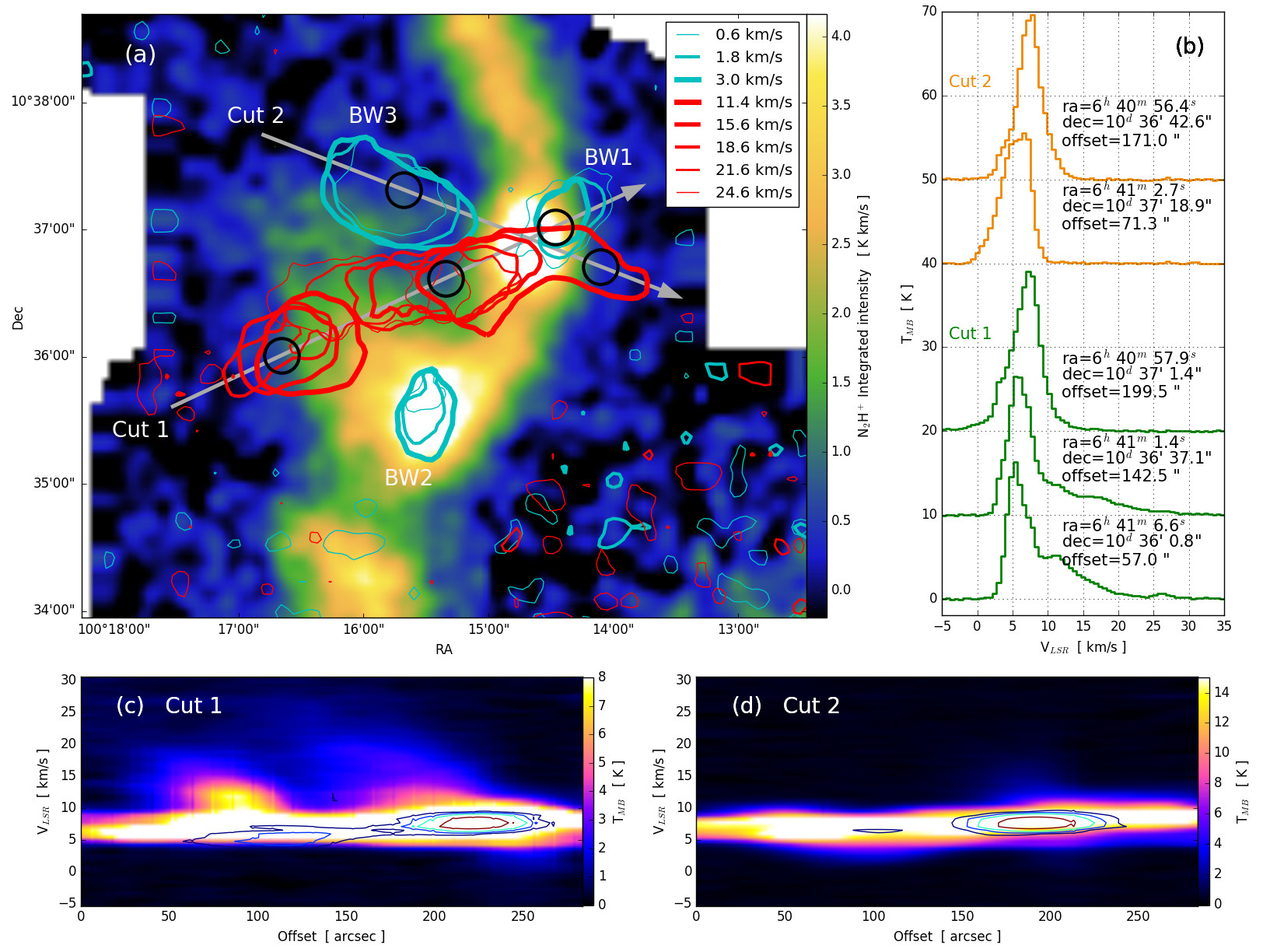} %Complete_figure_shock2_N2Hp.png}
      \caption{
      (a) Integrated map of \NtHp\, (in K km\,s$^{-1}$) of the junction region. The 
      contours show the emission of \CO\,(J=1-0) at large velocities with 
      respect to the bulk of the cloud. Each contour represent a different 
      velocity as indicated in the inserted legend, and the corresponding levels
      in \Tmb\, in order of increasing \Vlsr\, are 0.5 K, 1.0 K, 3.0 K, 
      3.0 K, 1.5 K, 1.1 K, 0.8 K and 0.6 K.
      (b) Spectra of \CO\,(J=1-0) emission at different locations along the cuts,
      averaged over the Gaussian beams represented by the black circles in 
      frame (a), whose diameters show the beam FWHM (18\arcsec).
      (c-d) Position-velocity diagrams of \CO\, emission along the grey arrows 
      shown in frame (a), with offset increasing from east to west. The colour 
      scales are cut to emphasise the fainter structures at large velocities. 
      The contours show the emission of the central lines of \NtHp\,J=1-0
      (unresolved blend of F$_1$F=21-11, 23-12 and 22-11 lines) at \Tmb=0.15, 
      0.3, 0.5, 1.0 K.}
      \label{fig:shock}
   \end{figure*}

   Our data enable us to characterise the outflows in the junction region with 
   a better spatial resolution than \citet{wolf-chase_star_2003}. Figure~\ref
   {fig:shock} shows the \CO(1-0) spectra in this region. The line profiles in
   frame (b) show that the bulk emission of the \CO(1-0) line peaks near 7 km\,s
   $^{-1}$, with a blue wing at $\Vlsr \sim 0-5$ km\,s$^{-1}$, at the 
   north-west edge of the source. When moving to the south-east, the bulk 
   emission peak shifts towards 5 km\,s$^{-1}$, while the blue wing disappears and a 
   strong red wing grows between $\Vlsr=10$ and 30 km\,s$^{-1}$.

   Frame (a) in Fig.~\ref{fig:shock} shows an \NtHp\, integrated intensity map 
   of the junction region with red and blue contours of \CO\, channel maps with 
   velocities of the red and blue line wings, respectively. The red-wing emission
   is distributed along an east-west elongated structure, in a direction 
   compatible with the flow towards HH 576. The blue wing contours are distributed
   in three blobs, hereafter referred to as BW1, BW2, and BW3 (Fig.~\ref{fig:shock} a),
   exhibiting a lack of bipolar symmetry with the red contours.

   Part of the asymmetry can be attributed to the large scale east-west
   velocity gradient. The bulk emission of source 1450, between -2 and +8 km\,s$^{-1}$, 
   makes the source appear in the contour plot despite the absence of blue wings.
   The suggestion by \citet{wolf-chase_star_2003} that BW2 might be an outflow 
   associated with their source 27S3 (s1450), is therefore irrelevant. 

   The location of BW1, at the north-west of source 1446, is mostly opposed to
   the elongated red-shifted emission on the south-east of source 1446. We 
   follow \citet{wolf-chase_star_2003} in interpreting them as a bipolar outflow 
   from the protostellar source 1446. Hereafter, we will refer to this outflow 
   as the main outflow. It was noted by \citet{reipurth_deep_2004} that the main 
   axis of this outflow is well aligned with the flow towards HH 576.

   The situation is more complex for BW3. It corresponds to no known dense sources
   in dust emission data. It presents a double peaked \CO\, line at $\sim 5$ and
   7 km\,s$^{-1}$, but the \tCO\, line shows a single peaked line at $\sim 7$ km\,s$^{-1}$, 
   suggesting that the red peak (7 km\,s$^{-1}$) traces the denser and less excited 
   matter, while the blue peak (5 km\,s$^{-1}$) arises from a more diffuse and excited
   component. \citet{wolf-chase_star_2003} proposed that this  blue-shifted
   emission is an outflow from their source 27S2 (s1448), whose red counterpart
   would be blended in velocity with the ambient matter, thus difficult to detect.
   However, (i) s1448 is not detected in \NtHp\,, an abundant species in 
   protostellar cores, (ii) the shape of the BW3 contours is elongated in the 
   direction of s1446, and (iii) the red contour at 11.4 km\,s$^{-1}$ extends to the
   south-west of the source 1446, in a direction opposed to BW3, making it an
   interesting candidate for the red counterpart of BW3. Hence, this latter red
   emission and BW3 could be the two components of a secondary outflow from 
   s1446. Overall, our data are compatible with the idea that both outflows 
   originate from the source 1446, which should therefore be a system of (at 
   least) two stars.

   The map in Fig.~\ref{fig:shock} shows that the red-shifted emission of the 
   main outflow is made of two parts: one $\sim 100\arcsec$ long part from s1446
   to the south-east, the other farther to south-east, spanning only $\sim 50
   \arcsec$. They are visible in the position-velocity (PV) diagram of Fig.~
   \ref{fig:shock} c, where they seem independent from each other, the first one
   being attached to s1446 (offset $\sim 200\arcsec$), disconnected of the second 
   one (offset $\sim 100\arcsec$). If the two parts trace the same outflow from 
   source 1446, the discontinuity in velocity between them remains to be explained.

   The velocity gradient within the first part of the main outflow shows
   velocities of $\sim 10$ km\,s$^{-1}$ near s1446, and $\sim 20$ km\,s$^{-1}$ at distances $\sim 
   100\arcsec$. Since the bulk velocity is $\sim 7$ km\,s$^{-1}$, it means that the more 
   distant gas flows faster, a fact in contradiction with the idea of a steady 
   outflow, and suggesting that the outflow is ejected in an intermittent fashion.
   In this scenario, all the material in the first part of the outflow would 
   have been ejected simultaneously, the faster parts of the gas reaching greater 
   distances. The velocities and distances are compatible with this scenario: the 
   gas traced by the contour at 24.6 km\,s$^{-1}$ lays at $\sim 100\arcsec$ from s1446, 
   while the contour at 11.4 km\,s$^{-1}$ traces gas typically within 25\arcsec of s1446.
   With a bulk velocity of 7 km\,s$^{-1}$ and assuming that the inclination of the velocity
   vector with the sightline is the same for both components, one finds the
   same ratio between the angular distance and radial velocity $\Delta \theta
   / \Delta \Vlsr \approx 5.7\arcsec/({\rm km}\,{\rm s}^{-1}$), that is the same age. In
   addition, the association with the HH 576 object 6\arcmin\,away (1.3 pc in 
   projection) already makes HH 576 a giant Herbig-Haro object \citep
   {reipurth_herbig-haro_2001, reipurth_giant_1997}. It is therefore unlikely 
   that the outflow main axis is too far from the plane of the sky. Assuming an 
   upper limit of 30 deg for this angle, the maximum radial velocities observed
   in the outflow of $\sim 30$ km\,s$^{-1}$ (23 km\,s$^{-1}$ relative to the source) correspond
   to $\sim 50$ km\,s$^{-1}$ relative to the source, and the travel duration of the 
   outflow is $\sim 10^4$ yr. This scenario also naturally explains the 
   discontinuity in velocity of the second part of the outflow if it was ejected
   during a previous outburst event. From its distance (125\arcsec) and its 
   radial velocity ($\sim 10$ km\,s$^{-1}$) relative to the source, one derives an age
   of $\sim 2 \times 10^4$ yr.

%   \subsubsection{Outflows from the north clump}

%   \rouge{Shock-like feature in CS in the north, with a compact-knot shape.
%   To be done.}

\subsection{Colliding filaments}
   \label{sec:colliding}

\subsubsection{Relative positions along the line of sight}
   \label{sec:geometry}

\begin{figure*}
      \includegraphics[width=0.5\textwidth]{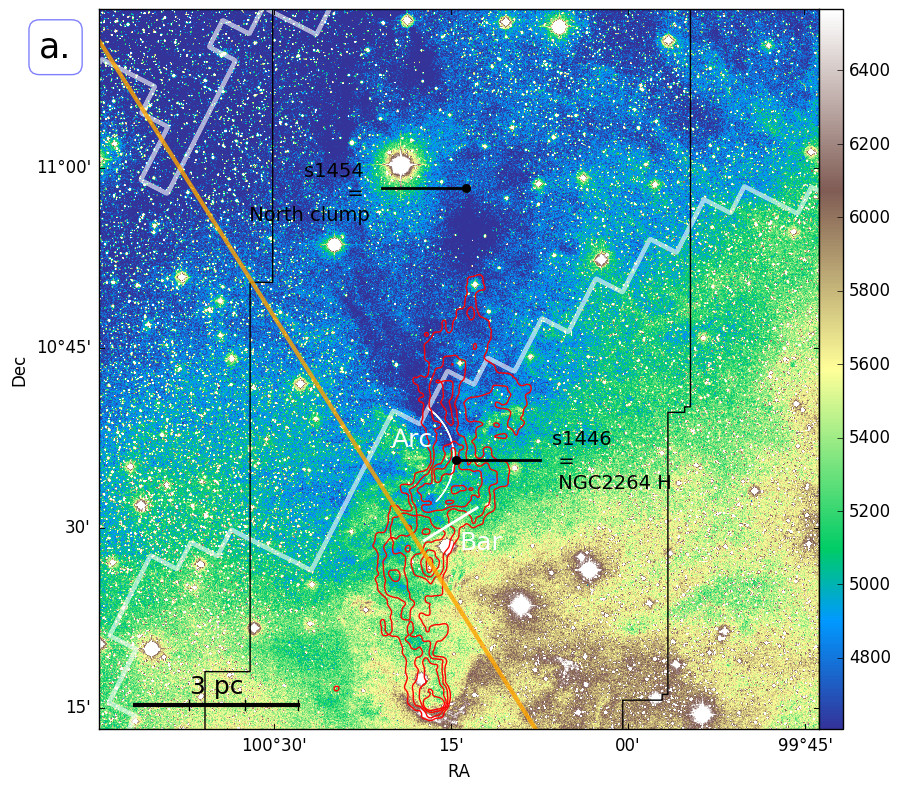}
      \hfill
      \includegraphics[width=0.5\textwidth]{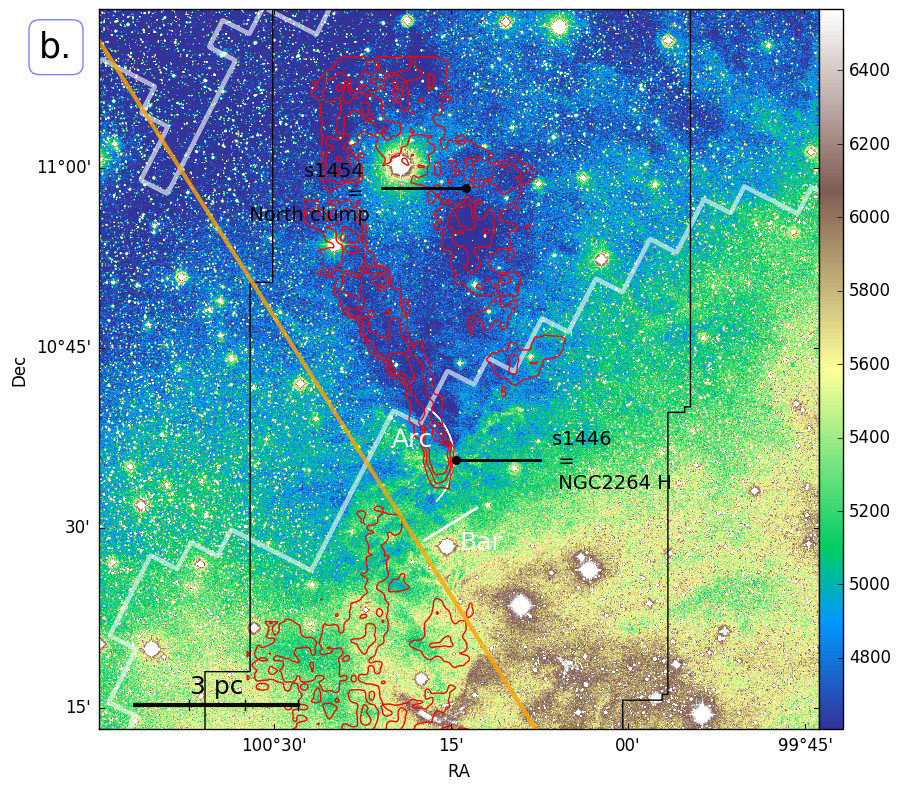}\\
      \includegraphics[width=0.5\textwidth]{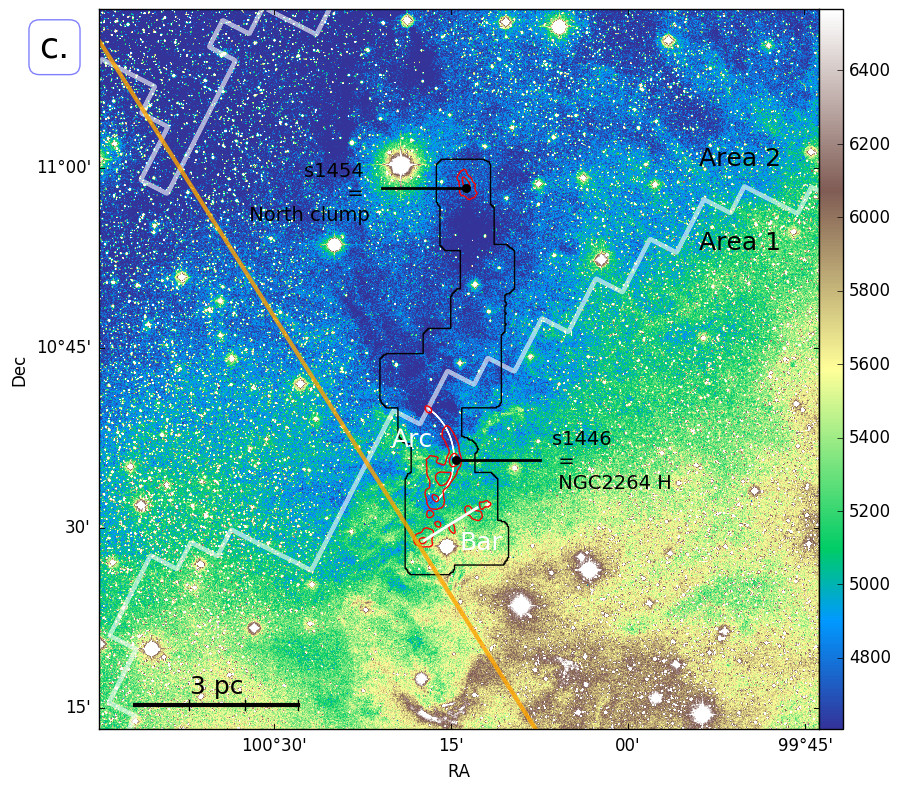}
      \hfill
      \includegraphics[width=0.5\textwidth]{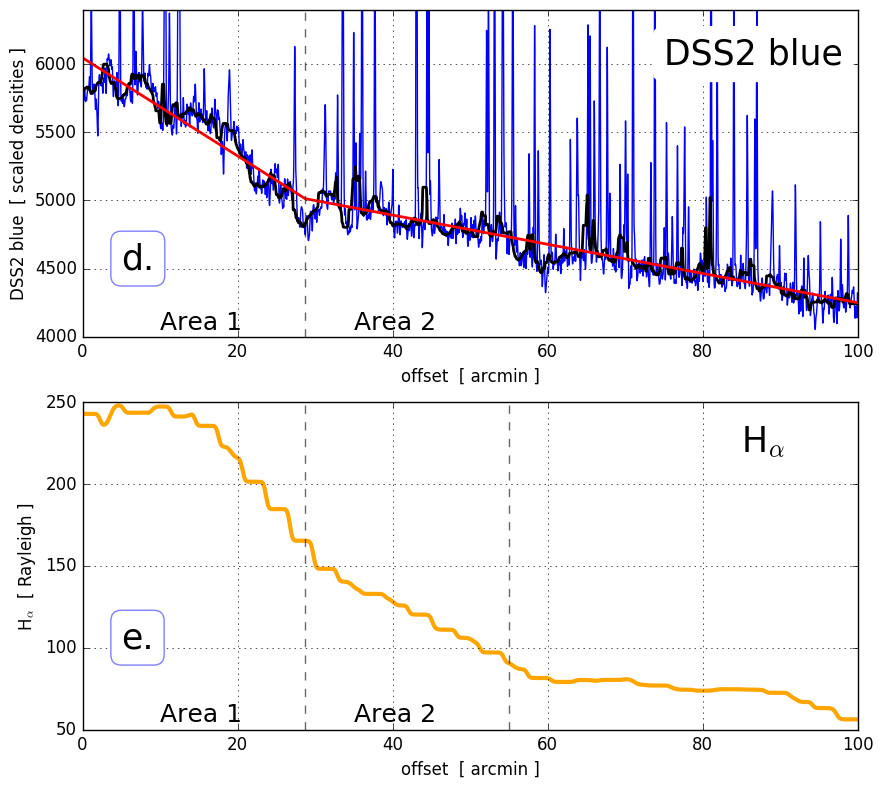}\\
      \caption{
      {\it a, b, c. } DSS2 blue map ($\lambda_{\rm eff} = 471$ nm) of G202.3+2.5. 
      The red contours show the \tCO\, emission from TRAO observations integrated
      between 6.3 and 9.3 km\,s$^{-1}$ ({\it a.}, with levels between $\int T_A^* dv =$ 
      3.3 and 8.25 K\,km\,s$^{-1}$ by steps of 1.65 K\,km\,s$^{-1}$) or 3.6 and 
      6.0 km\,s$^{-1}$ ( {\it b.}, with levels of $\int T_A^* dv=$ 2, 3 and 4 
      K\,km\,s$^{-1}$). In frame {\it c.} the contours show \NtHp, emission
      with levels of $\Tmb=$ 1 and 3 K. The thin black lines show the footprints
      of IRAM or TRAO observations. The thick white lines are isocontours of
      H$_\alpha$ emission at 150 and 80 Rayleighs. The arc and bar are shown
      with white lines. The sources 1446 and 1454 are shown in black. Frame {\it d.} shows
      the profile of DSS emission (blue line) along the cut between $(\alpha, \delta)$ 
      = (6:40:27.2, +10:11:37.4) and $(\alpha, \delta)$ = (6:43:53.9, 
      +11:31:28.8), whose intersection with the DSS map is shown by the orange
      line in frame {\it c}. The black line is the running median of the blue 
      curve, and the red line shows its two-piece linear best fit. Frame {\it e.}
      shows the profile in H$_\alpha$ along the same cut. The vertical dashed
      lines correspond approximately to the H$_\alpha$ contours in DSS maps.
      }
      \label{fig:DSS}
   \end{figure*}

\begin{figure}
      \includegraphics[width=0.5\textwidth]{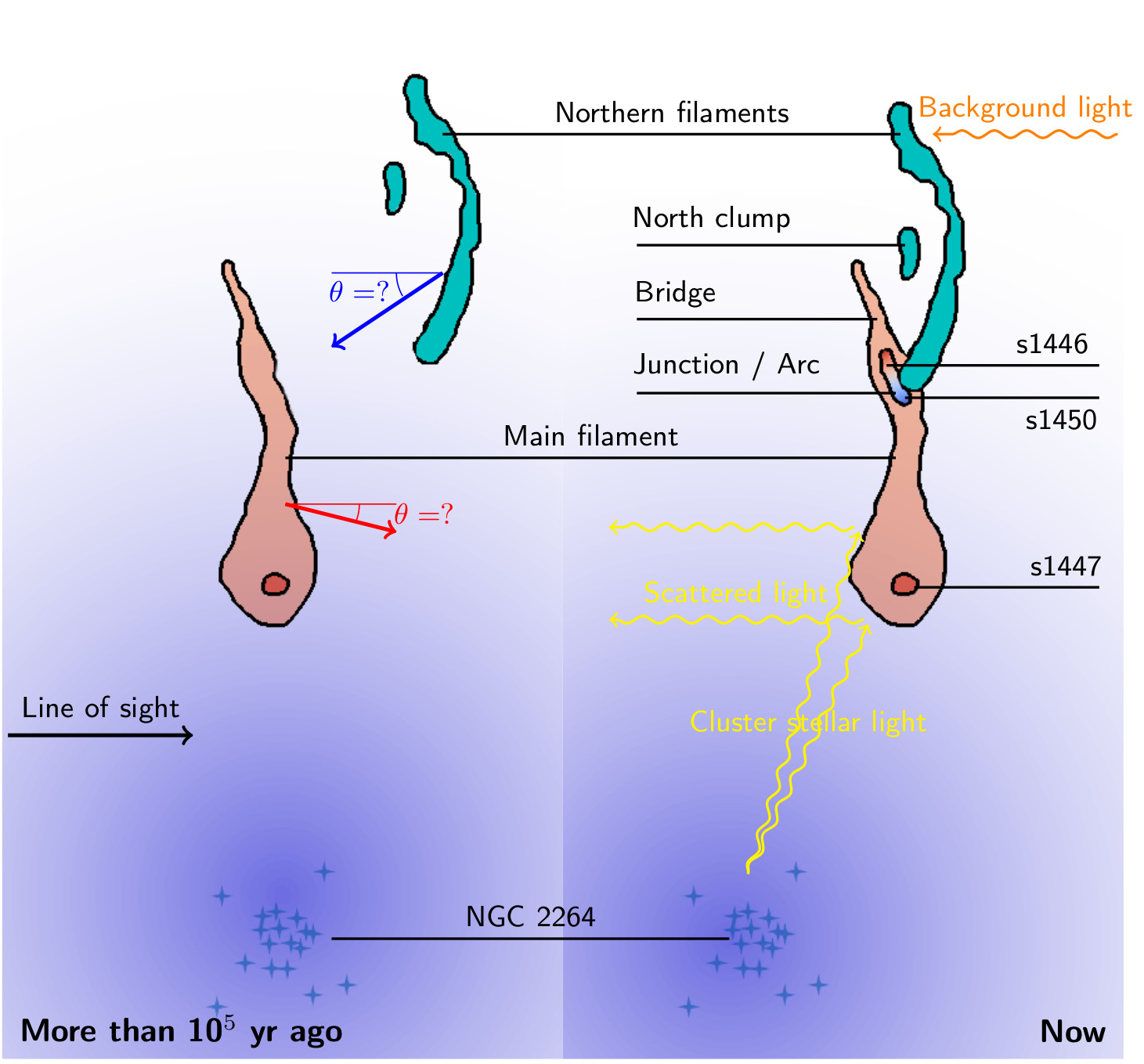}
      \caption{Sketch of the proposed geometry of G202.3+2.5. The red structure
      corresponds to the main filament, the blue ones to the northern filaments.
      The front side of the main filament is illuminated by the cluster stars, 
      and is responsible for most of the scattered light in the area 1 of 
      Fig.~\ref{fig:DSS}. The northern filaments are in the shadow of material
      between them and the cluster, and are seen in extinction in front of 
      background emission in the area 2 of Fig.~\ref{fig:DSS}.}
      \label{fig:sketch}
   \end{figure}

   The relative positions along the line of sight of the various velocity
   components and their position with respect to the rest of the Mon OB1 region
   cannot be determined solely from our millimetre observations. Extinction and
   scattering are better suited for this purpose. A $1^\circ\times 1^\circ$ DSS2 
   blue map ($\lambda_{\rm eff} = 471$ nm) of G202.3+2.5 is presented in 
   Fig.~\ref{fig:DSS}, showing a strong gradient in the average surface 
   brightness along the south-west to north-east direction.
   A surface brightness profile along this direction (Fig.~\ref{fig:DSS} d) 
   reveals that this gradient presents two different slopes. The steeper one is 
   in the area with values $\gtrsim 5000$ (arbitrary units), closer to the open 
   cluster NGC 2264, and corresponds to the H$_\alpha$ emission (hereafter 
   area 1, in blue in Fig.~\ref{fig:large_map}).
   It strongly suggests that the surface brightness is dominated by scattered
   light from the stars of NGC 2264 in this area, and by the field stars in the
   background and foreground of Mon OB1 in the rest of the field of view (hereafter
   area 2). This is summarised by a sketch in Fig.~\ref{fig:sketch}.

   Interestingly, the densest part of the junction region, traced by the \NtHp\,
   emission, is located just at the edge of area 1, but no structures seem to
   correspond in the DSS map. Instead, a relatively strong emission of scattered
   light with unrelated morphology is observed at this location. Because 
   extinction is strong at this wavelength, it implies that the structure 
   responsible for scattering is in front of the one traced by \NtHp.

   In contrast, the north-eastern and north-western filaments appear clearly in 
   the DSS map in extinction against the Galactic scattering background in area 
   2, forming structures which match very well the \tCO\, emission between 3.6 
   and 6 km\,s$^{-1}$. However, the southernmost tip of the north-eastern filament, 
   located next to the \NtHp\, emission, is not seen in extinction. This is 
   consistent with the idea that this \tCO\, emission and the \NtHp\, emission 
   arise from different layers of the same structure located behind the 
   scattering material.

   Figure~\ref{fig:DSS} also compares the morphology of the main filament, traced
   by the \tCO\, emission between 6.3 and 9.3 km\,s$^{-1}$, with the DSS map.
   The southernmost end of the main filament fits closely the inner part of a
   large arc seen both in scattered light in the DSS map and in mid-IR emission
   at 12 $\mu$m (Fig.~\ref{fig:large_map}), a typical tracer of photo-dissociation
   regions \citep[e.g. ][]{pilleri_evaporating_2012, anderson_wise_2014}. In the 
   first 15' northwards of the arc, the main filament is straight
   and oriented radially with respect to NGC 2264. In this part of the filament,
   there is a relatively good match between the crest of the filament and a
   decrease in the DSS surface brightness, however with a lesser contrast than
   in the case of the north-eastern and north-western filaments. Considering 
   that the column density in the main filament is $2-3$ times greater than in 
   the northern filaments, it cannot be in front of the scattering material. It
   cannot be behind it either, since a decrease in scattered light is observed.
   We conclude that the main filament is part of the scattering material. In addition,
   the shape of the arc, possibly seen edge-on or slightly from the back, and the 
   orientation of the filament suggest that its angle with the plane of the sky 
   is small.

   Gathering all these elements, we conclude that (i) the main filament is weakly
   inclined with respect to the plane of the sky and at a distance nearly equal
   to that of NGC 2264, (ii) it splits at the level of the junction region into 
   one short filament corresponding to the bridge, and one larger structure 
   stretching away from the observer, corresponding to the northern filaments.

   Finally, we note that the result presented in Sects.~\ref{sec:average_spectra} 
   and \ref{sec:gas_coldens} that the ratio between CO and \tCO\, brightness
   is larger in the main filament than in the north clump also suggests 
   different environment conditions in the two structures. This could be 
   naturally explained in the frame of our proposed geometry if the northern
   structures and the main filament used to be detached structures, and recently
   collided at the level of the junction region. We discuss further this idea in
   the following sections.

\subsubsection{Collision and inflow}
%\citet{gritschneder_oscillating_2017}

   Adopting the geometry obtained in the previous section, the radial velocities
   of the main filament (including the bridge) and of the northern filaments 
   imply that the two structures are moving towards each other with relative
   velocities of at least 2-4 km\,s$^{-1}$.

   This conclusion is strengthened by other elements. Figure~\ref{fig:NtHp_velo}
   shows the central velocity of the isolated hyperfine component of the \NtHp\,
   emission in the junction region. There is a clear velocity gradient from 
   $\sim 5$ km\,s$^{-1}$ in the south-east edge to $\sim 9$ km\,s$^{-1}$ in 
   the northern edge of the emission, so that this region, which is the densest
   of the cloud both in terms of column and volume densities, is also the one
   with the steepest velocity gradient. This can also be seen from \CeO\, emission
   in the right frame of Fig.~\ref{fig:PPV-centroids}. It is striking that the
   arc shape of the \NtHp\, emission is also followed by the integrated emission
   of \tCO\, between 6.3 and 9.3 km\,s$^{-1}$, corresponding to the main filament, 
   and seems to wrap around the integrated emission of \tCO\, between 3.6 and 6.0
   km\,s$^{-1}$, which corresponds to the northern filaments. Moreover the 
   lowest velocities in the \NtHp\, emission are reached where the two structures
   overlap. These observations strongly suggest that the dense arc traced by 
   \NtHp\, is the layer of gas compressed during the collision between the main
   filament and the southernmost tip of the northern filaments. The exact
   nature of the collision remains unclear, and two limiting scenarios can
   be considered, according to the relative values of the front and flow 
   velocities \citep[respectively perpendicular and parallel to the filament
   axis; see Fig. 1 in ][]{smith_nature_2016}.
   (i) The gas flows along the axis of the filaments which move slowly compared 
   to the flow velocity; in this case the collision region can be considered as 
   the convergence point of colliding flows, as reported, for example,
   by \citet{liu_top-scope_2018} in PGCC G26.53+0.17, or by \citet{peretto_sdc13_2014} 
   in the infrared dark cloud SDC13.
   (ii) The relative velocity of the filaments is significantly greater than
   the possible gas flow along them, making the process closer to a ballistic
   collision. Such a situation was reported, for example, by \citet
   {nakamura_cluster_2014} in Serpens south. Interestingly, in the hydrodynamical
   simulations by \citet{smith_nature_2016}, although the front velocities are
   generally larger than the flow ones, all types of filament collisions between
   our two limiting scenarios are observed.

   The shape of the \NtHp\, layer can be constrained by combining the total 
   column density from dust emission ($3-6 \times 10^{22}$ cm$^{-2}$, see Fig.~\ref
   {fig:map-Tdust-NH2}) and the volume density of this layer \citep[$\nH \approx 
   10^5$ cm$^{-3}$, ][]{lippok_gas-phase_2013}. If the total column density came
   solely from this layer, an upper limit of its length along the line of sight 
   can be derived and would be $\sim 0.2 - 0.4$ pc, a range of values marginally
   larger than the width of this layer in the plane of the sky. Therefore the
   region where \NtHp\, is observed is an arc-shaped filament, well represented
   by a bended cylinder of radius $\approx 0.1$ pc (along the line of sight and
   in the east-west direction) and length $\approx 1.3$ pc (in the south-north
   direction). With these dimensions and gas density, the \NtHp\, emitting gas
   corresponds to a mass of $141\,M_\odot$. As a comparison, the mass evaluated
   from the column density map of Fig.~\ref{fig:map-Tdust-NH2} within the area
   where the integrated \NtHp\, has a SNR$>5$ is 193 $M_\odot$, and in Table~\ref
   {tab:filament_masses} we reported a mass of 210 $M_\odot$ in the whole junction
   region for $\NHH>8 \times 10^{22}$ cm$^{-2}$. In principle it would leave 
   $\sim 50\,M_\odot$ for the envelop around the \NtHp\, arc, but the accuracy 
   of those mass estimates is not sufficient for the difference between these 
   numbers to be significant.

   This filamentary shape of the \NtHp\, emission suggests that if the hypothesis 
   of a collision is correct, it is more likely that it further compressed a 
   relatively dense and compact filament already present prior to the collision, 
   rather than created this dense structure from a diffuse and extended filament,
   in which case one would expect a more extended, sheet-like compression layer.
   This is supported by the fact that the \NtHp\, arc follows tightly the morphology
   of the main filament in Fig.~\ref{fig:NtHp_velo}, and also possibly
   by the typical chemical age of $\Delta t\sim 10^5$ yr reported by \citet
   {lippok_gas-phase_2013} for bright \NtHp\, emission. Indeed, if \nH\, has 
   increased during the collision by a factor $f$ only because of the decreasing 
   width $w$ along the collision direction, then the velocity of the collision 
   is $v_{\rm col} = f w / \Delta t$.
   The column density map shows values of $1-2 \times 10^{22}$ cm$^{-2}$ in
   the main filament, suggesting an increase by a factor of $f=3$ of the junction 
   region's density. The value of $w$ is between the length of the junction region 
   ($\sim 1$ pc), if the collision occurs mostly along the filament, as expected
   for the collision scenario (i), and its width ($\sim 0.2$ pc), in scenario (ii), 
   assuming that the collision occurs across the main filament. This leads to 
   $v_{\rm col} \sim 6 - 30 $ km\,s$^{-1}$, where the second value seems unlikely
   considering that there are no clear signs of a violent collision in our
   data \citep[e.g. no signs of shocked H$_2$ emission in WISE maps; see Fig.8 in][]
   {montillaud_galactic_2015}. Starting 
   from a diffuse structure would lead to even larger collision velocities and
   hence seems unlikely.
   %Since the 
   %observed difference in radial velocity is $\sim 4$ km\,s$^{-1}$, it would 
   %imply an angle of $\sim 40$ and 8 deg, respectively, between the collision 
   %axis and the plane of the sky.

   Interestingly this value of $10^5$ yr is an order of magnitude larger than 
   the age of the outflows out of the source 1446, as estimated in Paper I,
   suggesting that the protostar formed after the cloud collision started.

   Assuming a simple model where the collision occurs along a single direction
   with constant cross-section, one finds an inflow rate $\dot{M} = \frac{M}
   {\Delta t} (1-1/f)$. Using $f=3$ and $M=141$ M$_{\odot}$ leads to a rate of 
   mass inflow $\dot{M} \sim 1\times 10^{-3}\, M_\odot/{\rm yr}$ onto the initial
   filament, a value comparable to the infall rate found in massive clumps: 
   \citet{lopez-sepulcre_comparative_2010} found values between $10^{-3}$ and 
   $10^{-1}\,M_\odot/{\rm yr}$ in 48 high-mass clumps ($M\gtrsim 100\,M_\odot$);
   in a sample of 48 clumps, \citet{wyrowski_infall_2016} found $0.3 - 16 \times 
   10^{-3}\,M_\odot/{\rm yr}$ in nine other massive clumps; Traficante and coworkers
   also report infall rates in the range $10^{-5} - 2\times 10^{-3} {\rm M_\odot 
   yr}^{-1}$ in 16 70\,$\mu$m-quiet massive clumps \citep{traficante_massive_2017}
   and $0.7 - 46 \times 10^{-3} {\rm M_\odot yr}^{-1}$ in 21 massive clumps at different
   evolutionary stages \citep{traficante_testing_2018}. The fact that our estimate
   is of the same order of magnitude suggests that the collision process, with 
   the values proposed above, is a realistic scenario.

\subsection{Investigating the scenario of triggered star formation}

   As mentioned in the previous section, if the collision hypothesis is correct,
   the most likely scenario is that a pre-existing relatively dense filament had
   its density increased in the process of the collision. We showed that an
   increase by a factor of three is realistic. Since the observed mass of the \NtHp
   emitting gas is 141 $M_\odot$ for a length of 6\arcmin\, (1.3 pc), this structure
   has a linear mass of $M_{\rm lin} = 108\,M_\odot$/pc. This is in the
   range of values (92-186 $M_{\odot}$/pc) we found for the critical linear mass
   of gravitational instability under thermal and turbulent threshold in the 
   junction region (Sect.~\ref{sec:res_dust_coldens}). Considering that the \NtHp 
   emitting mass is only a lower limit of the mass estimate, it suggests that
   the structure is gravitationally unstable, consistently with the fact that 
   several protostars, including the most striking of the area (s1446), and 
   several starless cores are detected in this structure. The masses of these 
   % PS = {1446=21.6, 1448=15.3, 1451=16.3}, SL={1453=22.3, 1450=50.7}
   protostellar sources reported by \citet{montillaud_galactic_2015} add up to
   53.2 $M_\odot$, implying a high star formation efficiency of $\sim 20\%$ or 
   $\sim 38\%$ depending on which mass estimate is used for the \NtHp\,arc. In 
   contrast, prior to the collision, the original filament would have contained
   three times less gas, that is $M\sim 47\,M_\odot$, its linear mass would have been 
   $M_{\rm lin} \sim 36\,M_\odot$/pc, a value slightly below the critical linear
   mass of 37-92 $M_\odot$/pc found in the northern filaments (Sect.~\ref
   {sec:res_dust_coldens}), suggesting at best a moderate star formation 
   activity, comparable to that observed in the north-western filament.

   On the other hand, these masses include only the densest part of the junction
   region. The values in Table~\ref{tab:filament_masses} with a column density
   threshold of $8\times 10^{21}$ cm$^{-2}$ include larger areas with the 
   envelope of intermediate densities. Assuming that the $210 \,M_\odot$/pc 
   reported for the junction region contains $108\,M_\odot$/pc of dense gas, 
   if one divides only the mass of the dense layer by 3, the initial linear
   mass would have been $138\,M_\odot$/pc, a value similar to the one estimated
   for the southern part of the main filament.

   Altogether, it suggests that the initial filament was already forming stars
   and that the collision at least doubled its star formation activity. This
   situation is comparable to the one described by \citet{dale_dangers_2015},
   whose simulations demonstrate that triggering never consists in turning on
   star formation in an otherwise quiet cloud. Instead, it would imply the
   formation of stars that would not have formed without the triggering process,
   in a cloud which was already forming stars, so that both kinds of stars would
   be well mixed and extremely difficult to disentangle.

\subsection{The role of the environment}

   The fact that the junction region is located at the edge of the H$\alpha$ 
   emission, a tracer of \HII\, regions, suggests that the collision is related to 
   the possible expansion of the \HII\, region around NGC 2264. One possibility is
   that the main filament moves away from NGC 2264 as part of this expansion,
   and encountered the northern filaments. Alternatively, the northern filaments
   could be infalling towards NGC 2264, for example due to a global collapse of 
   the whole region, hence colliding the shell around the \HII\, region. It
   is therefore important to examine the age of the \HII\, region to determine
   whether it is still expanding or has reached an equilibrium with the 
   surrounding environment.

   The only ionising star in NGC 2264 is S Mon, a multiple stellar system which
   includes a O7V star and a O9.5V star with a total mass of $\sim 60\,M_\odot$
   \citep{gies_binary_1993, cvetkovic_new_2010}. We did not find an age estimate
   of S Mon itself, but \citet{sung_star_2011} reports the age estimates of the
   pre-main sequence stars in the same substructure as S Mon. The first method, 
   based on the SED fitting tool by \citet{robitaille_interpreting_2006}, leads
   to a median age of 1.5 - 2 Myr, with a dispersion between 0.2 and 3 Myr. The
   second method is based on isochrone fitting in colour-magnitude diagrams and
   leads to a median age of 3 Myr, with a dispersion between 1 and 6 Myr. \citet
   {venuti_gaia-eso_2018} find similar results using Gaia-ESO data. As these
   authors, we assume S Mon's age to be in the same range.

   \citet{williams_classical_2018} simulated the expansion of a spherical \HII\,
   region into a uniform cold and neutral environment. According to their model,
   for the ionising luminosity of S Mon \citep[$\log L_{EUV} = 48.7$ photons/s, ][]
   {venuti_gaia-eso_2018}, the maximum over-shoot radius of the \HII\, region is reached in
   $\sim 15$ Myr, and the equilibrium radius, slightly smaller than the maximum
   radius, is obtained after $\sim 40$ Myr. Hence, even with a conservative age
   estimate of 6 Myr, the \HII\, region is likely to still be expanding. With a 
   more realistic age of $\sim 2 - 3$ Myr, the ionising front and the shock
   fronts are still close to each other, with a radius of $\sim 70\%$ of the
   equilibrium radius. \citet{williams_classical_2018} quote $\sim 10$ pc as a 
   typical equilibrium radius, but there is a considerable scatter in this value
   \citep[e.g. $\sim 30$\,pc in $\lambda$ Ori,][]{liu_planck_2016}, and the complex
   geometry of the H$\alpha$ emission indicates that the environment was not
   uniform when the \HII\, region started its expansion. Therefore the distance of
   $\sim 12$ pc between S Mon and the edge of the H$\alpha$ emission in the 
   direction of the junction region seems a reasonable value for a still expanding
   \HII\, region. We conclude that the shell is still expanding, and that this 
   expansion is a possible candidate to explain the collision observed in the 
   junction region.

\begin{figure}
      \includegraphics[width=0.50\textwidth]{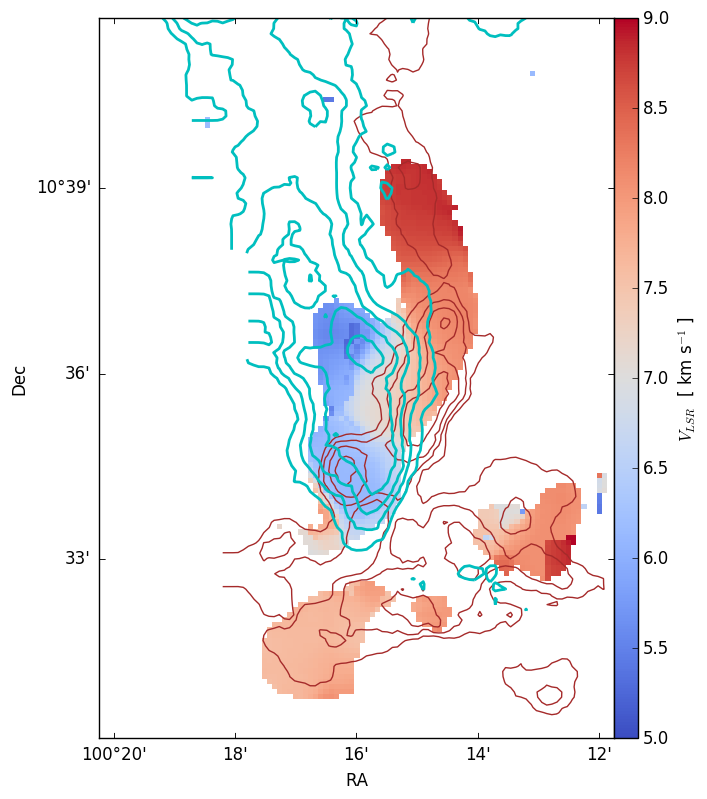}
      \caption{Centroid map of the isolated hyperfine component of \NtHp\,
      in the junction region. The light blue and dark red lines show the contours
      of the \tCO\, emission from the IRAM data integrated between 3.6 and 6.0
      km\,s$^{-1}$, and between 6.3 and 9.3 km\,s$^{-1}$, corresponding to the
      north-eastern filament and the main filament, respectively. The levels are
      between  3 and 18 K\,km\,s$^{-1}$ by steps of 1.8 K\,km\,s$^{-1}$, and 
      between 12 and 18 K\,km\,s$^{-1}$ by steps of 1.8 K\,km\,s$^{-1}$ for the 
      blue and red contours, respectively. In the white area, no line fitting was
      attempted because the \NtHp\, has a SNR<3.
      }
      \label{fig:NtHp_velo}
   \end{figure}

   On the other hand, it remains unclear whether the speed of the shock front 
   can be sufficient to explain the gap in velocities between the main filament
   and the northern filaments. It can be seen in Figs.~\ref{fig:res_velo_comp} 
   and \ref{fig:NtHp_velo} that the difference in radial velocities is $\sim 3-4$
   km\,s$^{-1}$, so that the complete velocity difference can be even larger,
   especially if the line between S Mon and G202.3+2.5 is nearly perpendicular to
   the line of sight. In their Fig.~1, \citet{williams_classical_2018} show a
   shock front with a velocity of only $\sim 1$ km\,s$^{-1}$. Therefore it
   remains a possibility that, in addition to the expansion, the northern 
   filaments have their own movement towards NGC 2264.

   A first glimpse to this possibility is provided by observing the large-scale
   velocity gradients within the Mon OB 1 molecular complex. The \CO\,(J=1-0) 
   data of the CfA 1.2m telescope survey by \citet{oliver_new_1996} show a
   velocity gradient from the region of the S Mon sub-cluster ($\Vlsr \approx 
   10-12$ km\,$^{-1}$) towards G202.3+2.5. A steeper gradient is also shown 
   in the south
   of the S Mon sub-cluster towards the Cone nebula which has radial velocities
   comparable to those of G202.3+2.5 ($\Vlsr \approx 4-8$ km\,$^{-1}$), but
   at a distance of only $\sim 5$ pc. Unfortunately the spatial and spectral 
   resolution of these data are not sufficient to fully investigate the movement
   of the northern filaments with respect to NGC 2264.
   The \CO\,(J=3-2) data by \citet{buckle_structure_2012}, obtained with the JCMT
   telescope reveal a wealth of details at 14\arcsec resolution, but for an area
   limited to $\sim 0.5^\circ$ around S Mon.

%=======================================================================================================================
\section{Conclusion and perspectives}\label{sec:conclusion}

We have studied G202.3+2.5, a complex filamentary, star-forming cloud at the edge
of the Monoceros OB 1 molecular complex, looking for a relationship between the
dynamics of the filament and its star formation activity. We have examined its 
general density and thermal structure using dust emission maps from the {\it Herschel}
Galactic cold cores programme, combined with TRAO 14-m and IRAM 30-m millimetre 
observations of mainly \CO, \tCO, \CeO, and \NtHp\, (J=1-0) molecular lines. From 
this data set, we also had characterised the compact sources in the
first paper of this series and examined the variations of their properties with
respect to the structure of the cloud. In addition here we characterised
the dynamics of the cloud from the radial velocity field derived 
from the millimetre data. These analyses were combined to draw a multi-scale view 
of G202.3+2.5, from core scales ($\sim 0.1$ pc) to the cloud scale ($\sim 
10$ pc), and led to propose a scenario of the recent star formation history of 
G202.3+2.5 in relation with its local environment in the Mon OB1 molecular complex.

The study has led to the following conclusions.
%\begin{itemize}
%   \item 
   G202.3+2.5 is found to be actively forming stars, and the filaments' linear
   masses suggest that this activity will be sustained in the near future.

%   \item 
   IRAM \CO\, observations confirm a known outflow from the brightest 
   and massive ($M_{\rm dust} \approx 22$ M$_\odot$) protostellar source 
   in G202.3+2.5, and suggest that this source could also be
   responsible for a second outflow in a different direction. The structure of
   the main outflow suggests that it was emitted in an intermittent fashion, and
   we estimated the oldest outburst, responsible for a large Herbig-Haro object,
   to be $\sim 2\times 10^4$ yr old.

%   \item 
   The structure derived solely from dust observations is very misleading. 
   The north-western filament, which looks like a single structure in {\it Herschel}
   maps, appears in IRAM data as the superposition of two independent velocity 
   components. In dust maps, the junction region seems to be the point where the
   main filament forks into two secondary filaments (the northern filaments), 
   whereas TRAO \tCO\, observations show that the two northern filaments are two
   parts of the same structure at $\Vlsr \sim 4$ km\,s$^{-1}$, while the 
   main filament is a different structure at $\Vlsr \sim 7-8$ km\,s$^{-1}$.

%   \item 
   The velocity components identified from the IRAM \tCO\, data show that
   the two main velocity structures, namely the main filament and the northern 
   filaments, merge in velocity in the junction region where the densest gas is 
   found and traced by \NtHp. The {\NtHp\,emission} exhibits a large velocity gradient,
   emitting continuously from the velocity of one structure to the other, 
   suggesting that the dense gas is a compressed layer between the two structures.

%   \item 
   The qualitative analysis of the extinction and scattered light in the
   visible DSS surface brightness map of the cloud led to the conclusion that
   the northern filaments are behind the main filament. Thus the shift in radial
   velocities implies that the two structures are moving towards each other, 
   supporting the idea of a collision between them in the junction region.

%   \item 
   Based on the typical chemical age of \NtHp\, in dense cores, we proposed
   a scenario in which the junction region would be the result of a collision 
   that started $\sim 10^5$ yr ago between the main filament and the southern tip
   of the northern filaments. We showed that in the frame of this scenario, our 
   data are compatible with the idea that the colliding part of the main filament 
   was already relatively dense and forming stars, similarly to the rest of this
   filament, and accreted additional gas from the northern filaments at a rate 
   $\dot{M} \sim 1 \times 10^{-3}\,M_\odot$/yr.

%   \item 
   We interpreted the local peak in star formation activity in the junction 
   region, and more specifically in the \NtHp\, emitting gas, as the result of 
   the increase in gas density, itself resulting from the collision. Therefore 
   this region falls under the frame of triggered star formation in the sense of 
   Type I triggering defined by \citet{dale_dangers_2015}: 'A temporary or 
   long-term increase in the star formation rate.'

%   \item 
   The collision scenario was put in perspective with the local environment
   of G202.3+2.5. The cloud is at the edge of the \HII\, region around the nearby 
   open cluster NGC 2264 and its most massive member, the O-type binary S Mon. 
   We showed that this \HII\, region is young enough to still be expanding, and
   is therefore a candidate for the origin of the collision. However, we cannot
   rule out the idea that the northern filaments move towards NGC 2264 as part
   of a possible global collapse of the region, contributing or even possibly 
   dominating the collision velocity.

%\end{itemize}

   The Monoceros OB 1 molecular cloud appears as a good target to characterise 
   the global evolution of a giant molecular cloud. Additional steps are
   already engaged to complement the multiscale approach of this study.

%\begin{itemize}

   %\item 
   The large scale \CO\, and \tCO\, (1-0) emissions in the Monoceros OB 1 
   molecular cloud will be observed to further study the connection between 
   G202.3+2.5 and NGC 2264 and test the ideas of global or hierarchical 
   collapse of this region.

   %\item 
   The collision in the junction region needs to be further characterised.
   Shock tracers, such as SiO(2-1) \citep{jimenez-serra_parsec-scale_2010} or
   H$_2$S and SO$_2$ transitions \citep{codella_shocked_2003}, will be observed
   to constrain the collision velocity and morphology. In addition, we will 
   look for infall signatures in the junction region, by searching asymmetry in 
   high-spectral resolution observations of HCO$^+$(1-0) in combination with the
   \NtHp(1-0) line to trace the systemic velocity \citep{fuller_circumstellar_2005}.

   %\item 
   Finally, the magnetic field geometry and strength, ideally measured at
   angular resolutions comparable to that of the molecular line tracers discussed 
   here, are key to understand the present state and the history of Mon OB 1. 
   We have measured dust polarized emission at 850 $\mu$m with the polarimetre 
   POL-2 at the JCMT 15-m telescope (Liu, Tie; et al., 2019, in preparation). 
   Such observations have proven to be good tracers of the plane-of-sky projected 
   B-field morphologies from large to small scales \citep[e.g. ][]{matthews_legacy_2009, 
   dotson_350_2010, hull_TADPOL:_2014, zhang_magnetic_2014, cortes_interferometric_2016, 
   koch_polarization_2018, liu_holistic_2018, liu_compressed_2018}. Given the 
   high densities in the junction regions, 
   the B-field morphology here could reveal dragged-in and pinched B-fields as 
   it is observed in gravity-dominated and collapsing cores \citep[e.g., ][]
   {girart_magnetic_2009, tang_evolution_2009-1}. In combination with molecular 
   line tracers, correlations between velocity gradients and B-field orientation
   can be searched to constrain the role of the B-field in the accretion process, 
   and the magnetic support can be evaluated to firmly conclude on core and 
   filament stability.

%\end{itemize}

\begin{acknowledgements}

   This work is based on observations carried out under project number 113-16 
   with the IRAM 30-m telescope. IRAM is supported by INSU/CNRS (France), MPG 
   (Germany) and IGN (Spain). JMo warmly thanks the staff of the 30-m for its
   kind and efficient help, and in particular Pablo Garcia for stimulating
   discussions.\\

   The project leading to this publication has received funding from the 'Soutien 
   à la recherche de l'observatoire' by the OSU THETA.\\

   This work was supported by the Programme National “Physique et Chimie du Milieu 
   Interstellaire” (PCMI) of CNRS/INSU with INC/INP co-funded by CEA and CNES.\\
   
   JMo, RB, DC, and VLT thank the French ministry of foreign affairs (French 
   embassy in Budapest) and the Hungarian national office for research and 
   innovation (NKFIH) for financial support (Balaton program 40470VL/ 
   2017-2.2.5-TÉT-FR-2017-00027).\\

   The project leading to this publication has received funding from the European 
   Union's Horizon 2020 research and innovation programme under grant agreement 
   No 730562 [RadioNet]\\

   This research used data from the Second Palomar Observatory Sky Survey (POSS-II),
   which was made by the California Institute of Technology with funds from the 
   National Science Foundation, the National Geographic Society, the Sloan 
   Foundation, the Samuel Oschin Foundation, and the Eastman Kodak Corporation.\\

   MJ and ERM acknowledge the support of the Academy of Finland Grant No. 285769.\\

   This research has made use of the SVO Filter Profile Service (http://svo2.cab.inta-csic.es/theory/fps/) 
   supported from the Spanish MINECO through grant AyA2014-55216.\\

   J.H.He is supported by the NSF of China under Grant Nos. 11873086 and 
   U1631237, partly by Yunnan province (2017HC018), and also partly by the 
   Chinese Academy of Sciences (CAS) through a grant to the CAS South America 
   Center for Astronomy (CASSACA) in Santiago, Chile.

   CWL was supported by the Basic Science Research Program through the National 
   Research Foundation of Korea (NRF) funded by the Ministry of Education, 
   Science and Technology (NRF-2019R1A2C1010851).
   
   SZ acknowledge the support of NAOJ ALMA Scientific Research Grant Number 2016-03B.

   KW acknowledges support by
   the National Key Research and Development Program of China (2017YFA0402702),
   the National Science Foundation of China (11973013, 11721303),
   and the starting grant at the Kavli Institute for Astronomy and Astrophysics, Peking University (7101502016).

\end{acknowledgements}

\bibliographystyle{aa}
%\bibliography{Paper_G202}
\bibliography{Article_G202}

\appendix

\section{Integrated intensity and channel maps}\label{anx:channelmaps}

\begin{figure}[h!]
   \includegraphics[height=0.35\textheight]{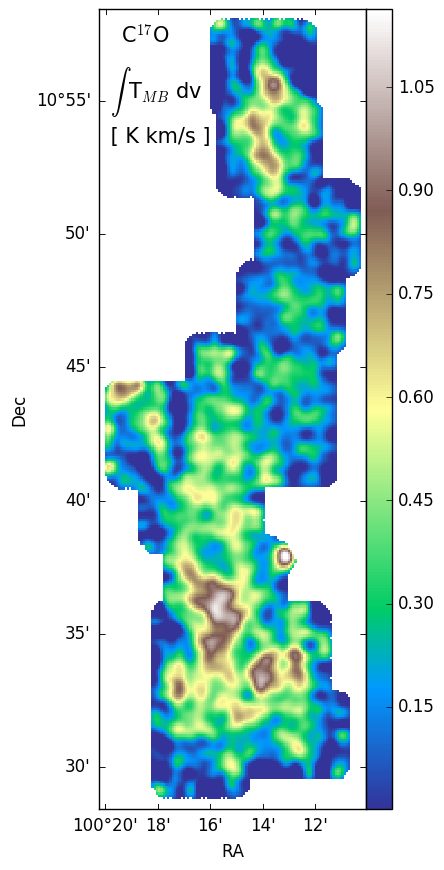}
   \hfill
   \includegraphics[height=0.35\textheight]{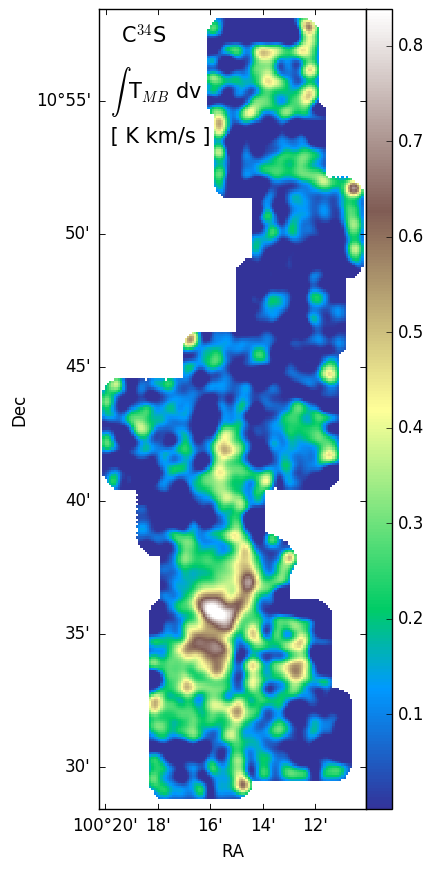}
   \caption{Integrated intensity maps of \CsO\, (left) and \CtS\, (right). The 
   maps are smoothed with a Gaussian kernel of FWHM=25\arcsec.}
   \label{fig:integmaps_other}
\end{figure}

\begin{figure*}
   \includegraphics[width=\textwidth]{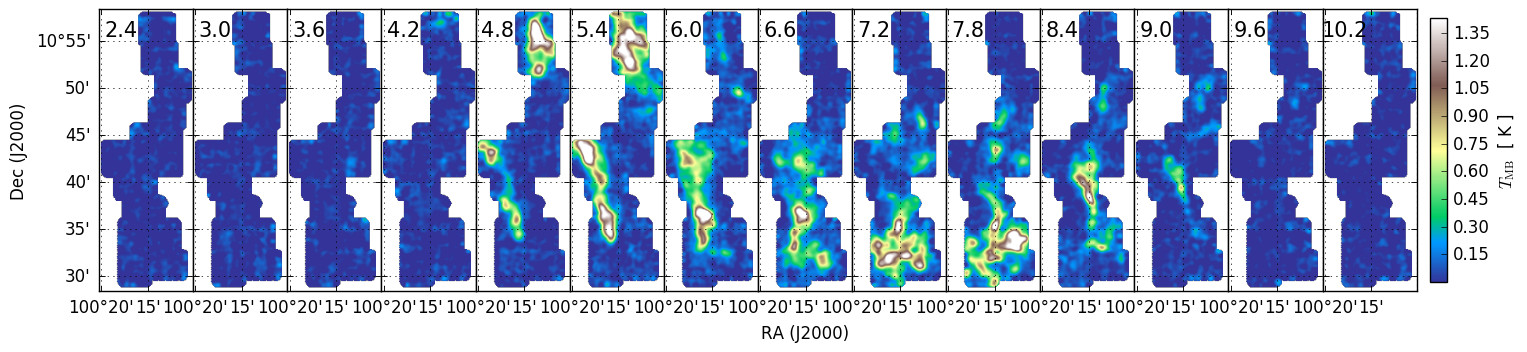}\\
   \includegraphics[width=\textwidth]{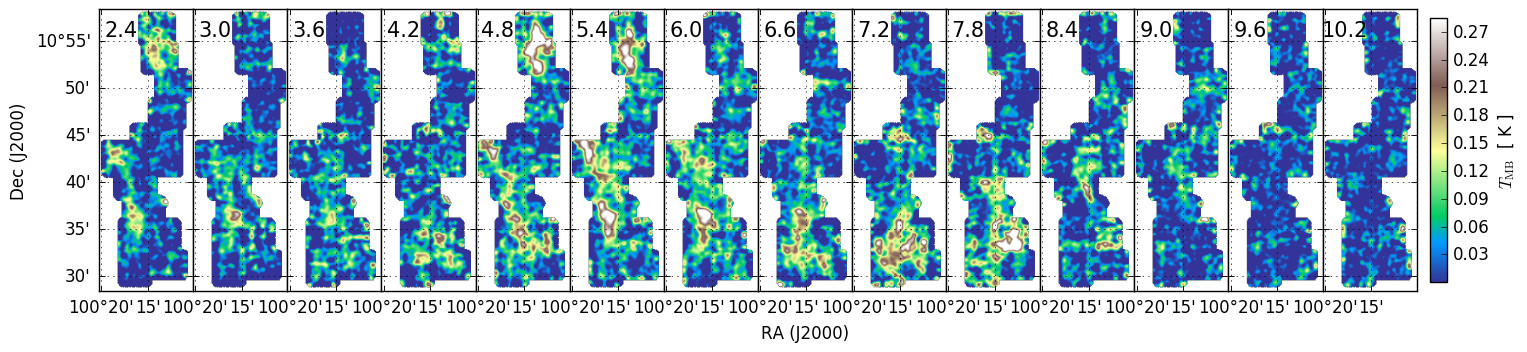}\\
   \includegraphics[width=\textwidth]{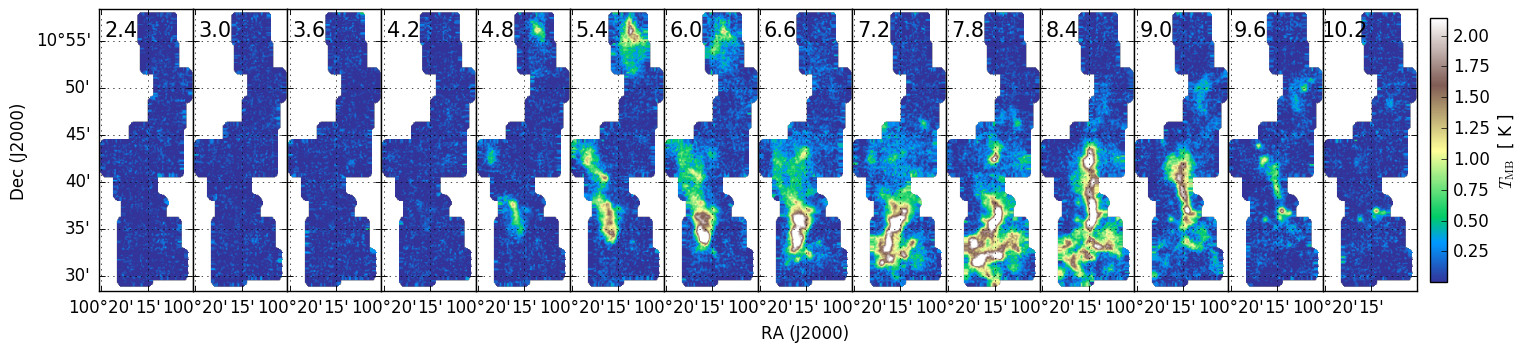}\\
   \includegraphics[width=\textwidth]{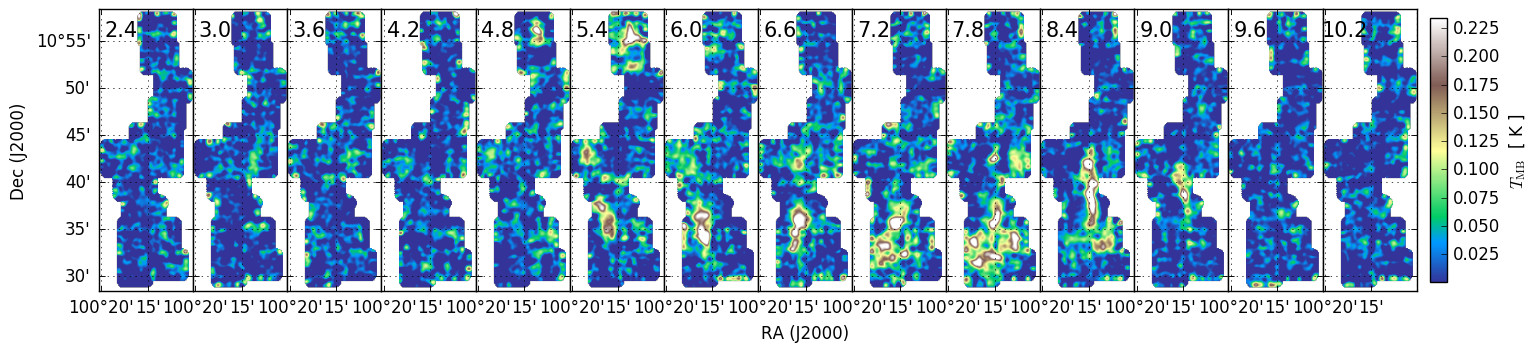}\\
   \includegraphics[width=\textwidth]{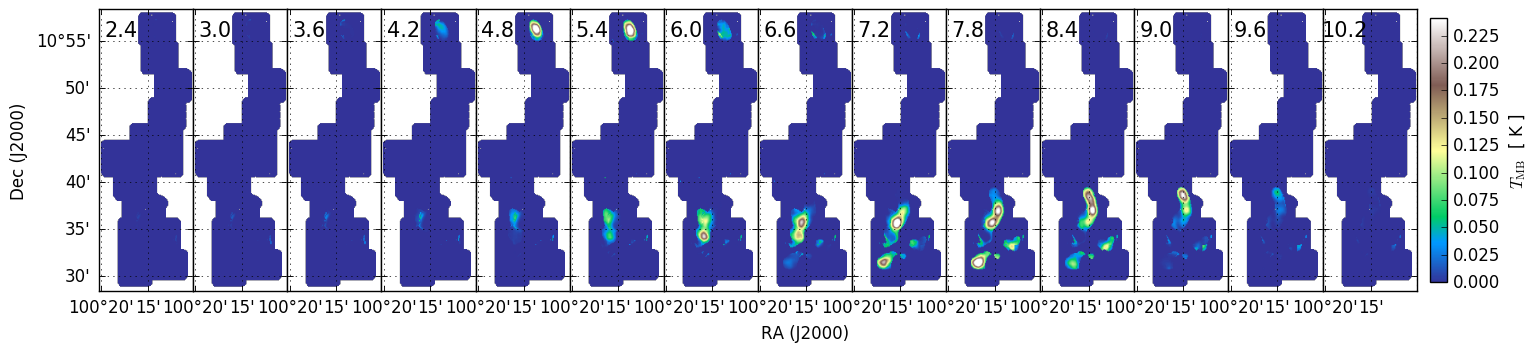}
   \caption{Channel maps of \CeO\,(top), \CsO\, (second row), CS (third row), 
   \CtS\, (forth row) and \NtHp\,(bottom). The maps of \CeO\, and \NtHp\, are
   smoothed with a Gaussian kernel of FWHM=5\arcsec, and \CsO\, and \CtS\, with
    FWHM=25\arcsec. For \NtHp, the hyperfine components were fitted assuming the
   LTE line ratios, and only the isolated component is shown here. For \CsO,
   the target line (F=3/2-5/2) is blended with the F=7/2-5/2 component, but the
   F=5/2-5/2 one is visible, shifted by $\Vlsr=-3.3$ km\,s$^{-1}$.}
   \label{fig:channelmaps_other}
\end{figure*}

\section{Linear mass formula for turbulent support}\label{anx:turbsupport}

   The critical linear mass of an isothermal, thermally and turbulence-supported
   cylinder is a usual quantity to characterise the stability of a
   self-gravitating interstellar filament. However, several versions of the
   formula are used in the literature that are mutually incompatible 
   \citep[e.g. ][]{liu_holistic_2018, pattle_jcmt_2017}, based on different
   recipes to convert between 1D and 3D velocity dispersions. We propose here a
   simple justification of the formulae used in Sect.~\ref{sec:res_dust_coldens},
   based on the pioneering works by \citet{chandrasekhar_fluctuations_1951} and 
   \citet{ostriker_equilibrium_1964}.

   \citet{ostriker_equilibrium_1964} demonstrated that the critical linear mass
   of a thermally supported isothermal infinite cylinder is:
   \begin{equation}
      M_{\rm lin} = \frac{2 k T}{\mu m G} = \frac{2 c_{\rm s}^2}{G}
      \label{eq:thsupport}
   \end{equation}
   where $T$ is the filament temperature, $m$ is the mass of the hydrogen nucleus,
   $\mu m$ is the mean particle mass, and $c_{\rm s} = \sqrt{2 k T / \mu m}$ is
   the so-called isothermal sound speed. We note that this quantity differs from 
   the actual sound speed by the adiabatic factor $\gamma$, and corresponds to
   the 1D rms of particle thermal velocity.

   \citet{chandrasekhar_fluctuations_1951} showed in the case of isotropic 
   microturbulence in a uniform medium that the equations by Jeans are modified
   by the transformation $c_{\rm s}^2 \rightarrow c_{\rm s}^2 + (1/3) U^2$, where 
   $U$ is the 3D rms turbulent velocity. We note that the factor $1/3$ makes the
   term $1/3 U^2$ correspond to the square of the 1D rms turbulent velocity.

   In Chandrasekhar's work, $c_{\rm s}$ is the actual (adiabatic) sound speed, 
   but the transformation can still be applied to Ostriker's formula since not
   only the difference with the isothermal sound speed is negligible compared to
   the uncertainties in our study, but also Chandrasekhar's assumption of an 
   adiabatic evolution of the gas is independent of the reasoning leading to the 
   transformation.

   From the observational point of view, the measured velocity dispersions are
   intrinsically 1D, along the line-of-sight. The thermal velocity line width
   corresponds to the isothermal sound speed, and we assume that the H$_2$ non-thermal
   velocity dispersion $\sigma_{\rm NT}$ is entirely caused by the isotropic 
   turbulence and therefore corresponds to the term $(1/3) U^2$ in Chandrasekhar's
   equations. As a result, the critical linear mass has the form:
   \begin{equation}
      M_{\rm lin} = \frac{2 (c_{\rm s}^2 + \sigma_{\rm NT}^2)}{G}
      \label{eq:turbsupport}
   \end{equation}
   where $c_{\rm s}$ can be estimated, for example, from dust temperature or from
   ammonia observations \citep{wilson_abundances_1994}, and $\sigma_{\rm NT}$
   can be derived from the gas temperature and the linewidth of a molecular
   tracer, for example \tCO\,(1-0) as in Sect.~\ref{sec:res_dust_coldens}.

   Finally, we note that it physically makes sense that only the 1D rms velocity
   dispersions appear in this equation. In the law of ideal gas, the thermal gas
   pressure is $P_{\rm th} = \rho kT/\mu m = \rho c_{\rm s}^2$, where $c_{\rm s}$
   is the 1D rms of particle thermal velocity. Similarly, the turbulent pressure
   is usually defined as $P_{\rm turb} = \rho \sigma_{\rm NT}^2$, where 
   $\sigma_{\rm NT}$ is also the 1D rms velocity \citep[between their Eq. 8 
   and 9]{mcKee_theory_2007}. Since Eq.~(\ref{eq:thsupport}) is equivalent to
   $M_{\rm lin} = 2 P_{\rm th} / \rho G$, it makes sense that the generalisation
   to turbulent support is $M_{\rm lin} = 2 (P_{\rm th}+P_{\rm turb}) / \rho G$,
   which is equivalent to Eq.~(\ref{eq:turbsupport}).
   
\end{document}